\documentclass[a4paper,fleqn,usenatbib]{mnras}

\usepackage[T1]{fontenc}
\usepackage{ae,aecompl}

\usepackage{graphicx}	
\usepackage{amsmath}	
\usepackage{amssymb}	
\usepackage{pdflscape}	
\usepackage{float}

\title[FIR emission of Seyfert galaxies]{The nuclear and integrated far-infrared emission of nearby Seyfert
galaxies\thanks{{\it Herschel} is an ESA space observatory with science instruments
provided by European-led Principal Investigator consortia and with important participation from NASA.}}

\author[Garc\'ia-Gonz\'alez et al.]{J. Garc\'ia-Gonz\'alez,$^1$
  A. Alonso-Herrero,$^{1, 2}$   
A. Hern\'an-Caballero,$^{1,3}$
\newauthor
M. Pereira-Santaella,$^4$
C. Ramos-Almeida,$^{5, 6}$
J. A. Acosta Pulido,$^{5, 6}$
T. D\'iaz-Santos,$^7$ 
\newauthor
P. Esquej,$^{8,9}$
O. Gonz\'alez-Mart\'in,$^10$
K. Ichikawa,$^{11}$
E. L\'opez-Rodr\'iguez,$^{12, 13}$
M. Povic,$^{14}$
\newauthor
P. F. Roche,$^2$ and
M. S\'anchez-Portal$^{8,9}$\\
\\
$^1$Instituto de F\'isica de Cantabria, CSIC-UC, Avenida de los Castros s/n, 39005 Santander, Spain\\
$^2$Department of Physics, University of Oxford, Oxford OX1 3RH, UK \\
$^3$ Departamento de Astrof\'isica y CC. de la Atm\'osfera, Facultad de CC. F\'isicas, Universidad Complutense de Madrid, E-28040 Madrid, Spain\\
$^4$Centro de Astrobiolog\'ia (CSIC/INTA), Ctra de Torrej\'on a Ajalvir, km 4, E-28850 Torrej\'on de Ardoz, Madrid, Spain\\
$^5$Instituto de Astrof\'isica de Canarias, E-38205, La Laguna\\
$^6$Departamento de Astrof\'isica, Universidad de La Laguna, E-38206 La Laguna\\
$^7$N\'ucleo de Astronom\'ia de la Facultad de Ingenier\'ia, Universidad Diego Portales, Av. Ej\'ercito Libertador 441, Santiago, Chile\\
$^8$European Space Astronomy Centre (ESAC)/ESA, P.O. Box 78, 28690 Villanueva de la Ca\~{n}ada, Madrid, Spain\\
$^9$ISDEFE, Beatriz de Bobadilla 3, 28040 Madrid, Spain\\
$^10$Centro de Radioastronom\'ia y Astrof\'isica (IRyA-UNAM), 3-72 (Xangari), 8701 Morelia, Mexico\\
$^{11}$National Astronomical Observatory of Japan, 2-21-1 Osawa, Mitaka, Tokyo 181-8588, Japan\\
$^{12}$Department of Physics \& Astronomy, University of Texas at San Antonio, One UTSA Circle, San Antonio, TX 78249, USA \\
$^{13}$Department of Astronomy, University of Texas at Austin, 1 University Station C1400, Austin, TX 78712, USA \\
$^{14}$Instituto de Astrof\'isica de Andaluc\'ia (IAA-CSIC), E-18008 Granada, Spain\\
}

\date{Accepted XXX. Received YYY; in original form ZZZ}

\pubyear{2015}

\begin{document}
\label{firstpage}
\pagerange{\pageref{firstpage}--\pageref{lastpage}}
\maketitle

\begin{abstract}
 We  present far-infrared (FIR) $70-500\,\mu$m imaging observations obtained with 
{\it Herschel}/PACS and SPIRE of 33 nearby (median distance of 30 Mpc) Seyfert galaxies from the Revised
Shapley-Ames (RSA) catalogue. We obtain the FIR nuclear ($r=1\,$kpc and $r=2\,$kpc)  
and integrated spectral energy distributions (SEDs). We 
estimate the unresolved nuclear emission at 70 $\mu$m and
 we fit the nuclear and integrated FIR SEDs with a grey body model.
 We find that the integrated FIR emission
of the RSA Seyferts  in our sample is dominated by emission from the host galaxy, with dust properties similar to those
of normal galaxies (non AGN). We use four criteria to select galaxies whose nuclear $70\,\mu$m emission
has a significant AGN contribution: (1) elevated 70/160 $\mu$m
flux ratios, (2)spatially resolved, high   dust 
temperature gradient, (3) $70\,\mu$m excess emission with respect to the 
fit of the FIR SEDs with a grey body,   and (4)  excess of nuclear SFR obtained from 
$70\,\mu$m over SFR from mid-infrared indicators.  16 galaxies (48 per cent of the initial sample) satisfy at least one of these conditions,
whereas  10 satisfy half or more. After careful examination of these, 
we select  six {\it bona fide} candidates
 (18 per cent of the initial sample) 
and estimate that $\sim 40-70$ per cent of their nuclear ($r=1-2\,$kpc) $70\,\mu$m emission 
is contributed by dust heated by the AGN.  
\end{abstract}


\section{Introduction}
\label{Section-introduction}
 
Seyfert galaxies emit a large fraction
  of their bolometric emission in the far-infrared (FIR)
  range \citep{Rieke1978}.  This emission is due to dust grains that re-radiate the absorbed optical and
 ultraviolet photons emitted by the active galactic nucleus (AGN) as well as  
 by stars in their host galaxies.
 \cite{Rodriguez-espinosa} studied the FIR emission of a sample of
 optically selected Seyfert galaxies using observations taken with the 
 {\it Infrared Astronomical Satellite (IRAS)}. They found that
   the distribution of  the integrated FIR luminosities and $60\,\mu$m to $100\,\mu$m colours
   of Seyfert and starburst galaxies were indistinguishable and concluded that most
 of the FIR emission of Seyfert galaxies must be due to star formation processes.
 Using {\it Infrared Space Observatory (ISO)} observations, \cite{Perez-garcia} showed that the
  integrated FIR thermal emission of Seyferts can be modelled with a combination 
 of three different components: warm, cold, and very cold dust. The warm component is due to
 dust heated by the AGN and/or 
 circumnuclear star formation (T$\sim150\,$K), the cold component  comes from dust heated by stars in the
 disc of the galaxy (T$\sim 40-50\,$K) and the very cold component 
 arises from dust heated by the general interstellar radiation field
 of the galaxy (T$\sim10-20\,$K).  \cite{Spinoglio2002} also found using {\it ISO}
 imaging data that the integrated FIR emission of
   Seyfert galaxies is dominated by emission from the host galaxy.
 
 \cite{Mushotzky} used
 FIR observations taken with the Photodetector Array Camera (PACS; \citealt{Poglitsch}) 
on board {\it Herschel} to study a sample of hard X-ray selected 
 galaxies from the 58 month {\it Swift} Burst Alert
Telescope (BAT) Active Galactic Nuclei catalogue. They found that
$>$ 35 per cent and 20 per cent of the sources are point-like at 70 and 160 $\mu$m, respectively.
Using the same sample, \cite{Melendez} showed that 
the  integrated FIR luminosity distributions of Seyfert 1 and Seyfert 2 galaxies
are similar and their integrated $f_\nu(70\mu{\rm m})/f_\nu(160\mu{\rm m})$ ratios
are indistinguishable  from those of normal galaxies.  \cite{Hatziminaoglou2010}
studied 469 spectroscopically confirmed AGN. They 
used the Spectral and Photometric Imaging REceiver
 (SPIRE; \citealt{Griffin}) data and showed that the FIR emission of
Seyfert 1 and Seyfert 2 galaxies is
identical to that of star forming galaxies. The AGN contributes
very little to the integrated FIR emission and  its contribution becomes important at wavelengths shorter
than 70 $\mu$m, where the torus starts playing an important role.

 According to the unified model of AGN \citep{Antonucci}, there is a dusty torus
 surrounding the accreting supermassive black hole.  Dusty torus models
 where the dust is distributed both homogeneously (e.g., \citealt{Fritz}) or in a clumpy
 configuration  (see e.g., \citealt{Nenkova2008,Hönig-kishimoto}) predict emission in the
 FIR, although the torus emission peaks in the mid-infrared (MIR) spectral range. Based on
 studies of individual Seyfert galaxies using {\it Herschel}
 observations,  the contribution of the dust heated by the AGN 
 to the total FIR emission varies from galaxy to galaxy. 
For example, \cite{Almudena} found that in NGC~1365
 the AGN is the brightest  source in the MIR but
 does not dominate in the FIR. Using  the \cite{Nenkova2008} torus models they quantified the 
 AGN emission at $70\,\mu$m and determined that the AGN only contributes
 at most 1 per cent within the central $5.4\,$kpc. For NGC~2992, \cite{Ismael} showed
 that the AGN dominates  the emission between 15 and 30 $\mu$m, but its contribution decreases
 rapidly for wavelengths $>30\,\mu$m. \cite{Cristina} studied
 NGC~3081 and found that the FIR nuclear luminosity within a radius of $\leq$ 0.85 kpc
 was well reproduced with the \cite{Nenkova2008} clumpy torus models and concluded that
 the AGN dominates  the FIR nuclear luminosity of this galaxy.
  There are  however no statistical studies of the AGN contribution in the FIR, so 
 it is important to find a method to determine if the AGN dominates in the 
 FIR for large samples of galaxies.
  \cite{Mullaney2011} studied the infrared emission of a sample of local X-ray
selected AGN with little evidence of host galaxy contamination
in their MIR {\it Spitzer}/IRS spectra. They found that at least 3 of the 11 AGN in their sample
are AGN dominated even at 60 $\micron$.

 In this work we study the FIR ($70-500\,\mu$m) emission of a sample of 33 nearby
 (median distance of 30\,Mpc) Seyfert galaxies 
 drawn from the Revised Shapley-Ames catalogue (RSA; \citealt{Sandage-Tammann})
 using {\it Herschel} imaging observations taken with PACS and
 SPIRE.  The main goal is to disentangle the
 FIR emission of these Seyfert galaxies due to dust heated by the AGN from that due to
 dust heated by star formation. In particular,  
 we take advantage of the {\it Herschel} angular resolution of 5.6\,arcsec 
 at $70\,\mu$m, which provides a median physical resolution of 
  $0.8\,$kpc for our sample of galaxies. This allows us to 
 study the nuclear (radii of $r = 1\,$kpc and $r=2\,$kpc) and integrated FIR emission
 of Seyfert galaxies. The paper is organized as follows.  In Section~\ref{Section-sample}
 we present our  sample selection and  the comparison with the  entire RSA sample. In
 Section~\ref{Section-observations} we describe the data reduction and
 derive the aperture photometry. Section~\ref{Section-results} presents our 
 results, such as the unresolved $70\,\mu$m emission, the FIR colours, 
 the grey body fitting and the nuclear and extranuclear star
 formation rates (SFR). In Section~\ref{Section-criteria}  we put forward a number
   of criteria to  identify those 
Seyfert galaxies
 in our sample whose $70\,\mu$m emission is mostly due 
 to dust heated by the AGN  and discuss the {\it bona fide} candidates.
 The conclusions are presented in Section \ref{Section-conclusions}.
  Throughout this work we use a cosmology with $H_0=73\,{\rm km\,s}^{-1}~{\rm Mpc}^{-1}$,
 $\Omega_m=0.27$ and $\Omega_\Lambda=0.73$.

\section{The sample}
\label{Section-sample}

 We selected a sample of 33 nearby (distances $D_L<70\,$Mpc, Table~\ref{galaxy-sample})
  Seyfert galaxies \citep[see][]{Maiolino-Rieke} from the RSA
  catalogue (\citealt{Sandage-Tammann})  with  
  {\it Herschel}/PACS imaging observations in at least two bands and SPIRE imaging observations
  from our own programs and from the archive\footnote{http://www.cosmos.esa.int/web/herschel/science-archive}
  (see Table~\ref{table-observing-programs}). We imposed the distance
criterion so we could obtain at least one nuclear (radii of $r = 1\,$kpc  and/or $r=2\,$kpc) 
FIR measurement at $70\,\mu$m. We also required Seyfert galaxies with existing high angular
resolution ($0.3-0.4\,$\arcsec) MIR spectroscopy 
\citep{Hoenig2010,GonzalezMartin2013,Esquej2014,AAH2015}
obtained on $8-10\,$m class telescopes (T-ReCS, CanariCam  and VISIR instruments).
We used the spectra published in these references instead of reducing
the archival data.
These observations allow us to determine whether they have  star formation activity
on typical physical scales of $50-60\,$pc, which is necessary
when trying to determine what galaxies in our sample
have AGN-dominated FIR emission.  23 of these galaxies also have
estimates of the nuclear and integrated SFR from mid-IR {\it Spitzer}/IRS spectroscopy
 taken from \cite{Diamond-Stanic-Rieke}  (hereafter DSR2012).

In Table~\ref{galaxy-sample} we list the properties of the Seyfert galaxies in our sample
including their luminosity distance, optical apparent 
magnitude ($B_T$), morphological
type,  and the  optical activity type  (15 Sy 1 galaxies and 18 Sy 2 galaxies).
 We consider as Sy 1 the 1.5 and 1.9 Seyfert galaxies.

  We obtained the {\it Spitzer}/IRS SL+LL spectra from the Cornell
  Atlas of Spitzer/IRS Sources  version 7 \citep[CASSIS,][]{Lebouteiller2011,Lebouteiller2015}.
   The stitching of the 
  different spectral orders has been made as described in 
  \cite{Antonio2016}. We also provide in Table~\ref{galaxy-sample} the
equivalent width (EW) of
the 6.2 and 11.3 $\mu$m PAH features measured from {\it Spitzer}/IRS
short-low (SL) spectra, except for NGC~1068 which
was from a short-high (SH) spectrum \citep[see ][]{Esquej2014}.
 We measured the EW of the PAH features
following the method described by \cite{HernanCaballero2011}.
 Finally, we give the AGN bolometric luminosity, and
the  nuclear ($r=1\,$kpc) and extranuclear ($r>1\,$kpc) SFRs taken from
 DSR2012.  For those RSA Seyferts not in  that
  work we take the AGN bolometric luminosities from \cite{Mason2012} and
 \cite{Esquej2014}.

\begin{landscape}
\begin{table}
 \caption{Galaxy sample.}
 \label{galaxy-sample}
    \begin{tabular}{@{}cccccccccccc}
  \hline
  \tiny Number& \tiny Name & \tiny $D_L$   & \tiny  ${B_T}^1$ & \tiny  Morphological$^2$ 
  & \tiny   Activity  & \tiny  EW of 6.2$\mu$m  & \tiny 
  EW of 11.3$\mu$m  & \tiny log ${L_{bol,AGN}}^6$& \tiny SFR (r=1kpc)$^6$& \tiny SFR (r$>$1kpc)$^6$ & \tiny Ref. Activity\\
  & \tiny & \tiny  (Mpc)  & \tiny (mag)  & \tiny   type& \tiny  type & \tiny  PAH ($\mu$m)& \tiny  PAH ($\mu$m)& \tiny ($erg~s^{-1}$)
  & \tiny ($M_{\sun} yr^{-1}$)& \tiny ($M_{\sun} yr^{-1}$)& \tiny type \\
  \hline

  \tiny 1& \tiny  ESO~323-G077 & \tiny  60.2  & \tiny  13.58 & \tiny  (R)SAB0\textasciicircum 0(rs) & \tiny  Sy 1.2 & \tiny 0.049$\pm$0.005& \tiny  0.126$\pm$0.005 & \tiny 43.9$^{8}$ & \tiny & \tiny &\tiny 3\\
  \tiny 2& \tiny  IC~5063 & \tiny  49.9  & \tiny  13.22 & \tiny  SA0\textasciicircum +(s)? & \tiny   Sy 2 & \tiny $<$0.018 & \tiny 0.011$\pm$0.004 & \tiny 44.0$^{8}$ & \tiny & \tiny  & \tiny 10 \\
  \tiny 3& \tiny  Mrk~1066 & \tiny  49.0 & \tiny  13.64 & \tiny  (R)SB0\textasciicircum +(s) & \tiny  Sy 2 & \tiny 0.544$\pm$0.007  & \tiny 0.591$\pm$0.005& \tiny  & \tiny  & \tiny   & \tiny 4 \\
  \tiny 4& \tiny  NGC~1068 & \tiny  14.4 & \tiny  9.61 & \tiny  (R)SA(rs)b & \tiny  Sy 2 & \tiny   & \tiny  & \tiny 44.3$^{8}$ & \tiny & \tiny & \tiny 5 \\
  \tiny 5& \tiny  NGC~1320 & \tiny  35.5 & \tiny  13.32 & \tiny  Sa? edge-on & \tiny  Sy 2 & \tiny   & \tiny  & \tiny  & \tiny & \tiny  & \tiny 3 \\
  \tiny 6& \tiny  NGC~1365 & \tiny  21.5 & \tiny  10.21 & \tiny  SB(s)b & \tiny  Sy 1.8 & \tiny  0.258$\pm$0.004  & \tiny 0.314$\pm$0.002 & \tiny 44.3 & \tiny  4.80 & \tiny  8.40 & \tiny 3 \\
  \tiny 7& \tiny  NGC~1386 & \tiny  10.6 & \tiny  12.00 & \tiny  SB0\textasciicircum +(s) & \tiny   Sy 2 & \tiny $<$0.019 & \tiny 0.072$\pm$0.005 & \tiny 43.5 & \tiny  0.05 & \tiny  & \tiny 11 \\
  \tiny 8& \tiny  NGC~1808 & \tiny  12.3 & \tiny  10.76$^2$ & \tiny  (R)SAB(s)a & \tiny  Sy 2 & \tiny 1.078$\pm$0.006 & \tiny 1.013$\pm$0.003 & \tiny 41.2$^{8}$ & \tiny & \tiny  & \tiny 7 \\
  \tiny 9& \tiny  NGC~2110 & \tiny  32.4 & \tiny  14.00 & \tiny  SAB0\textasciicircum - & \tiny   Sy 2 & \tiny 0.014$\pm$0.007  & \tiny 0.051$\pm$0.005 & \tiny 43.7$^{8}$& \tiny & \tiny  & \tiny 12 \\
  \tiny 10& \tiny  NGC~2273 & \tiny 28.7 & \tiny  12.55 & \tiny  SB(r)a? & \tiny  Sy 2 & \tiny 0.273$\pm$0.006   & \tiny 0.383$\pm$0.007 & \tiny 43.6 & \tiny 0.76& \tiny   & \tiny 4 \\
  \tiny 11& \tiny  NGC~2992 & \tiny 34.1 & \tiny  12.80 & \tiny  Sa pec & \tiny   Sy 1.9 & \tiny 0.295$\pm$0.008 & \tiny 0.328$\pm$0.017  & \tiny 44.7 & \tiny   0.77 & \tiny  0.54  & \tiny 1 \\
  \tiny 12& \tiny  NGC~3081 & \tiny 34.2 & \tiny  12.68 & \tiny  (R)SAB0/a(r) & \tiny   Sy 2 & \tiny 0.022$\pm$0.012 & \tiny 0.046$\pm$0.005 & \tiny 44.6 & \tiny   0.15 & \tiny  0.31  & \tiny 13 \\
  \tiny 13& \tiny  NGC~3227 & \tiny  20.6 & \tiny  11.55 & \tiny  SAB(s)a pec & \tiny  Sy 1.5 & \tiny 0.215$\pm$0.006  & \tiny 0.359$\pm$0.007 & \tiny 44.0 & \tiny  0.48 & \tiny  0.21 & \tiny 3 \\
  \tiny 14& \tiny  NGC~3281 & \tiny  44.7 & \tiny  12.62 & \tiny  SA(s)ab pec? & \tiny  Sy 2 & \tiny 0.013$\pm$0.008  & \tiny 0.010$\pm$0.011 & \tiny 45.0 & \tiny  0.87 & \tiny  & \tiny 3 \\
  \tiny 15& \tiny  NGC~3783 & \tiny  36.1 & \tiny  12.89 & \tiny  (R')SB(r)ab & \tiny  Sy 1.5& \tiny 0.001$\pm$0.005 & \tiny 0.013$\pm$0.009  & \tiny 44.2 & \tiny   0.05 & \tiny  & \tiny 3 \\
  \tiny 16& \tiny  NGC~4051 & \tiny  12.9 & \tiny  10.93 & \tiny  SAB(rs)bc & \tiny   Sy 1.5 & \tiny 0.089$\pm$0.004  & \tiny 0.114$\pm$0.003 & \tiny 43.5 & \tiny  0.13 & \tiny  0.88& \tiny 1 \\
  \tiny 17& \tiny  NGC~4151 & \tiny  20.3 & \tiny  11.13 & \tiny  (R')SAB(rs)ab? & \tiny  Sy 1.5 & \tiny 0.005$\pm$0.003 & \tiny 0.013$\pm$0.003  & \tiny 44.5 & \tiny  0.06 & \tiny  & \tiny 3 \\
  \tiny 18& \tiny  NGC~4253 & \tiny  61.3 & \tiny  13.30 & \tiny  (R')SB(s)a? & \tiny   Sy 1.5 & \tiny 0.087$\pm$0.005 & \tiny 0.083$\pm$0.005 & \tiny  & \tiny  & \tiny   & \tiny 1 \\
  \tiny 19& \tiny  NGC~4258 & \tiny  7.98 & \tiny  8.95 & \tiny  SAB(s)bc & \tiny   Sy 1.9& \tiny 0.034$\pm$0.013 & \tiny 0.067$\pm$0.009 & \tiny 42.2 & \tiny  0.05 & \tiny  0.39& \tiny 1 \\
  \tiny 20& \tiny  NGC~4388 & \tiny  17.0 & \tiny  11.83 & \tiny  SA(s)b? edge-on & \tiny  Sy 2 & \tiny 0.072$\pm$0.005 & \tiny 0.140$\pm$0.003 & \tiny 44.4 & \tiny  0.21 & \tiny  0.24  & \tiny 14 \\
  \tiny 21& \tiny  NGC~4507 & \tiny  59.6 & \tiny  12.81 & \tiny  (R')SAB(rs)b & \tiny   Sy 2 & \tiny 0.019$\pm$0.006  & \tiny 0.049$\pm$0.003 & \tiny 44.6 & \tiny   0.99 & \tiny& \tiny 10\\
  \tiny 22& \tiny  NGC~4579 & \tiny  17.0 & \tiny  10.56 & \tiny  SAB(rs)b & \tiny   Sy 1.9& \tiny   & \tiny & \tiny 42.4$^{9}$  & \tiny  0.05 & \tiny  0.47& \tiny 1 \\
  \tiny 23& \tiny  NGC~4594 & \tiny  12.7 & \tiny  9.28 & \tiny  SA(s)a edge-on & \tiny   Sy 1.9 & \tiny   & \tiny  & \tiny 42.5  & \tiny  0.03 & \tiny  0.66& \tiny 1 \\
  \tiny 24& \tiny  NGC~4725 & \tiny  27.0 & \tiny  10.11$^2$ & \tiny  SAB(r)ab pec & \tiny  Sy 2 & \tiny   & \tiny & \tiny 41.9 & \tiny  0.01 & \tiny  0.34 & \tiny 3 \\
  \tiny 25& \tiny  NGC~5135 & \tiny  57.7 & \tiny  12.94 & \tiny  SB(s)ab & \tiny  Sy 2 & \tiny  0.742$\pm$0.009 & \tiny 0.777$\pm$0.007  & \tiny 44.9 & \tiny   6.10 & \tiny  3.60& \tiny 3 \\
  \tiny 26& \tiny  NGC~5347 & \tiny  40.2 & \tiny  13.40 & \tiny  (R')SB(rs)ab & \tiny  Sy 2 & \tiny  0.046$\pm$0.001 & \tiny 0.059$\pm$0.005 & \tiny 43.5$^{8}$ & \tiny   & \tiny & \tiny 3 \\
  \tiny 27& \tiny  NGC~5506 & \tiny  30.0 & \tiny  12.79 & \tiny  Sa pec edge-on & \tiny   Sy 1.9& \tiny 0.012$\pm$0.004 & \tiny 0.055$\pm$0.003 & \tiny 44.8 & \tiny 0.58 & \tiny   & \tiny 1 \\
  \tiny 28& \tiny  NGC~7130 & \tiny  68.7 & \tiny  12.98$^2$ & \tiny  Sa pec & \tiny  Sy 1.9 & \tiny  0.416$\pm$0.011  & \tiny 0.434$\pm$0.008 & \tiny 44.3 & \tiny 4.30 & \tiny  6.70 & \tiny  3\\
  \tiny 29& \tiny  NGC~7172 & \tiny  37.6 & \tiny  12.85 & \tiny  Sa pec edge-on & \tiny  Sy 2 & \tiny  0.052$\pm$0.006  & \tiny 0.205$\pm$0.009 & \tiny 44.2 & \tiny 0.79 & \tiny  0.68 & \tiny 3 \\
  \tiny 30& \tiny  NGC~7213 & \tiny  24.9 & \tiny  11.18 & \tiny  SA(s)a? & \tiny  Sy 1.5& \tiny  $<$0.025 & \tiny 0.059$\pm$0.006 & \tiny 43.1$^{8}$ & \tiny  0.04 & \tiny  0.39  & \tiny 1 \\
  \tiny 31& \tiny  NGC~7465 & \tiny  28.4 & \tiny  13.31 & \tiny  (R')SB0\textasciicircum 0?(s) & \tiny   Sy 2& \tiny & \tiny   & \tiny  & \tiny  & \tiny   & \tiny 15 \\
  \tiny 32& \tiny  NGC~7479 & \tiny  32.4 & \tiny  11.70 & \tiny  SB(s)c & \tiny  Sy 1.9 & \tiny     & \tiny 0.008$\pm$0.008 & \tiny 43.2 & \tiny 0.32 & \tiny  1.70& \tiny 3 \\
  \tiny 33& \tiny  NGC~7582 & \tiny  22.0 & \tiny  11.46 & \tiny  (R')SB(s)ab & \tiny   Sy 2 & \tiny 0.508$\pm$0.033 & \tiny 0.703$\pm$0.011  & \tiny 44.5 & \tiny 2.10 & \tiny  1.90 & \tiny 1 \\

  \hline
  
      \multicolumn{12}{l}{\tiny $^1$ from \citet{Maiolino-Rieke};  $^2$ NED Homogenized from \citet{Vaucouleurs}
      ;$^3$  from \citet{Veron-Veron}; $^4$ from \citet{Contini};}\\
      \multicolumn{12}{l}{\tiny  $^5$ from \citet{Osterbrock-Martel}; $^6$ from \citet{Diamond-Stanic-Rieke};}\\
      \multicolumn{12}{l}{\tiny    $^{7}$ from \citet{Brightman-Nandra}; $^{8}$ from \citet{Esquej2014};
      $^{9}$ from \citet{Mason2012}; $^{10}$ from \citet{Kewley2001}; $^{11}$ from \citet{Reunanen};}\\
      \multicolumn{12}{l}{\tiny $^{12}$ from \citet{Bradt}; $^{13}$ from \citet{Phillips}; $^{14}$ from \citet{Trippe};
      $^{15}$ from \citet{Malizia}}\\
      
    \end{tabular}

\end{table}
\end{landscape}

\begin{figure*}
    \begin{center}
      
      \includegraphics[width=0.45\textwidth]{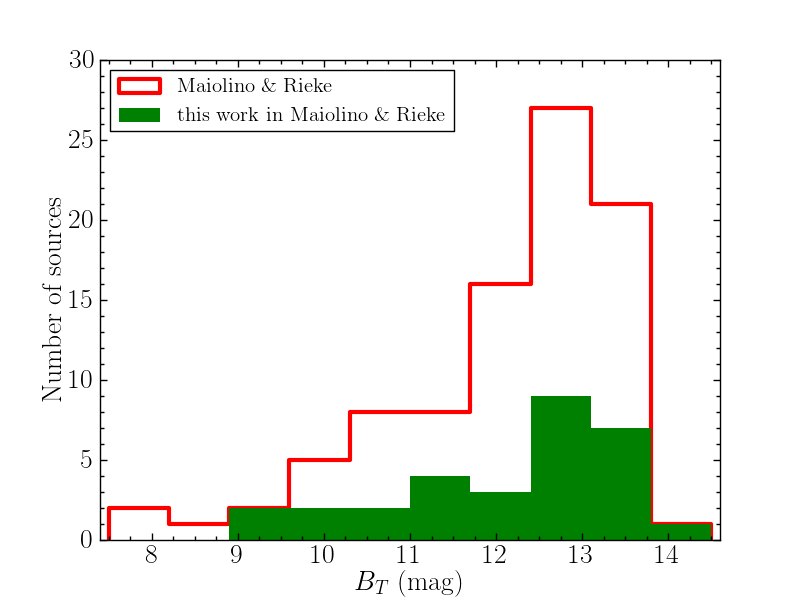}{\vspace{0cm}}
      \includegraphics[width=0.45\textwidth]{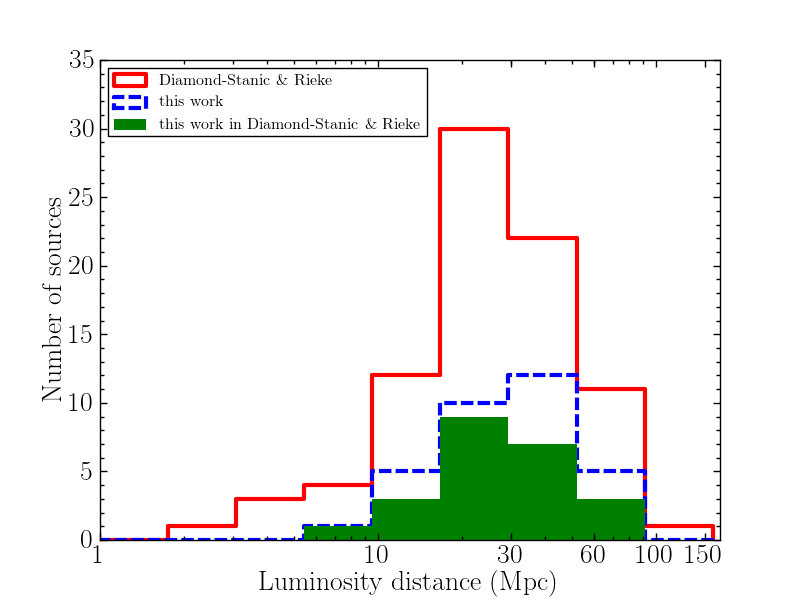}{\vspace{0cm}}
      \includegraphics[width=0.45\textwidth]{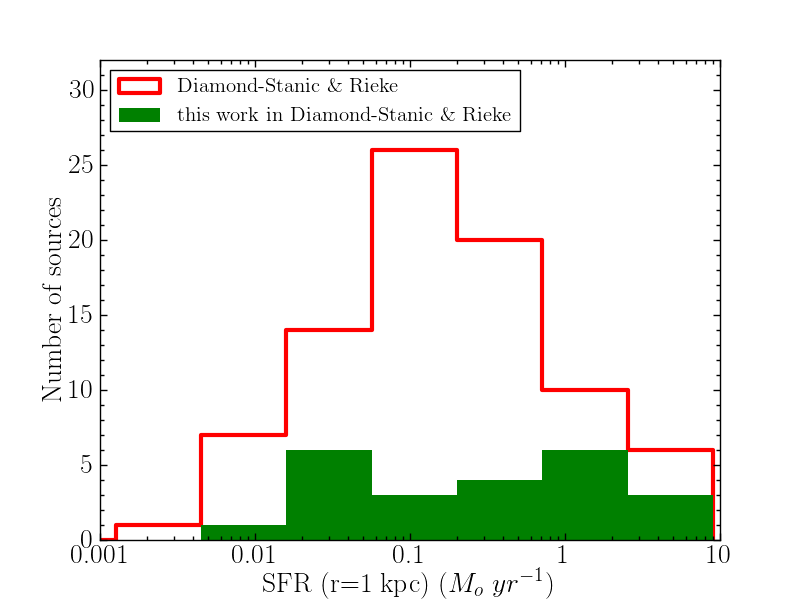}{\vspace{0cm}}
      \includegraphics[width=0.45\textwidth]{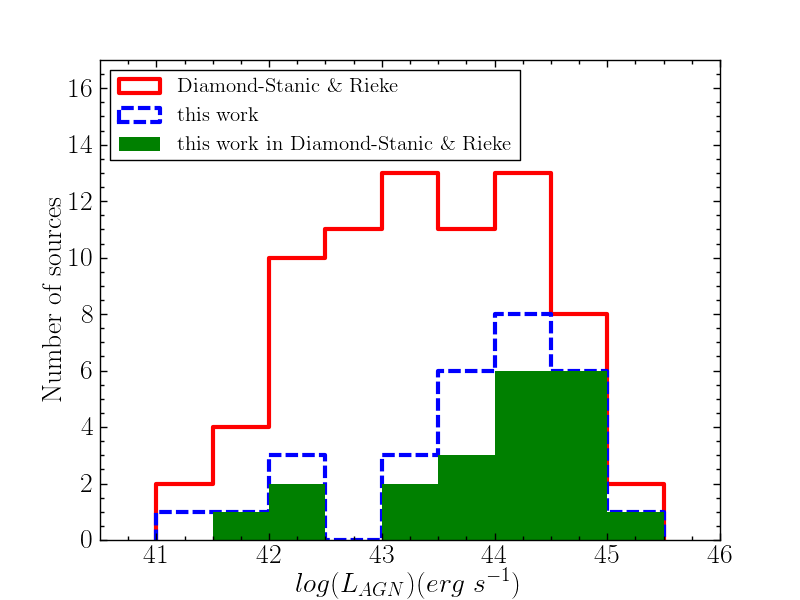}{\vspace{0cm}}

  \caption{Sample comparison. Top left panel: Distribution of 
  the optical apparent magnitude, $B_T$, for the RSA galaxies
  in \citet{Maiolino-Rieke} (91 galaxies, in red)
 and our sample  (30 galaxies, in green). 
  Top right panel: Distribution of the luminosity distance for the galaxies in  DSR2012
  (84 galaxies, in red) and our sample
  (33 galaxies, in  dashed blue). There are 10 galaxies in our sample not included in the
   DSR2012 sample. The galaxies in common 
  between  DSR2012 sample and our sample are shown in green.
  Bottom left panel: Distribution of the nuclear SFR (r=1kpc) for the galaxies in  DSR2012
  (84 galaxies, in red) and our sample with
   DSR2012 SFR values (23 galaxies, in  green).
  Bottom right panel: Distribution of the AGN bolometric
  luminosity for the galaxies in  DSR2012 
  (74 galaxies, in red), our sample (29 galaxies, in  dashed blue) and our sample with
   DSR2012 values (21 galaxies, in blue).} 
   
  \label{sample-comparison}

  \end{center}
\end{figure*}

\subsection{Sample comparison}

Since the RSA sample is selected based on the galaxy optical apparent magnitude, we used the 
 $B_T$ values from \citet{Maiolino-Rieke} to determine if our galaxy selection is 
representative of the entire RSA  sample of Seyfert galaxies.
Their sample contains 91 relatively nearby Seyfert galaxies
with $B_T<$13.31 and have 30 galaxies in common with our sample.
The top left panel of Fig.~\ref{sample-comparison} shows the $B_T$ distribution 
for the two samples. The 91 galaxies in 
\citet{Maiolino-Rieke} are shown in red and the 30 of 33 galaxies in our sample are shown in blue.
       Inspection of  Fig.~\ref{sample-comparison} and a Kolmogorov-Smirnov (K-S)
      test show that our sample is not significantly different from the RSA sample in terms of $B_T$ 
      (p-value$=0.92$).

\begin{table*}
\begin{minipage}{180mm}
 \caption{Summary of the statistical properties of our sample and comparison samples.}
 \label{tabla-comparacion-muestras}
 \begin{tabular}{lcccccccc}
  \hline
  &\multicolumn{4}{c}{RSA sample}& \multicolumn{4}{c}{This work}\\
 Quantity&Number& Mean& $\sigma$ & Median &Number & Mean& $\sigma$ & Median\\
  \hline
  
  $B_T$ $^1$  &91& 12.08 & 1.32 & 12.55 & 30 & 12.10 & 1.34 & 12.68\\
  Luminosity distance (Mpc) $^2$  & 84&  30.5 & 21.1 & 24.4 & 33 (23)* & 32.2 (29.5)* & 16.4 (15.6)* & 30.0 (27.0)*\\
  SFR (r=1kpc) ($\rm M_{\sun} yr^{-1}$) $^2$ & 84& 0.67 & 1.42 & 0.15 & 23 & 1.03 & 1.66 & 0.32\\
  log ${L_{AGN}}~(\rm erg~s^{-1})$ $^2$ & 74& 43.4 & 1.0 & 43.4& 29 (21)* & 43.8 (44.0)* &  1.0 (0.9)* & 44.0 (44.3)*\\
  \hline
  
  \end{tabular} 
  
  $^1$  From \citet{Maiolino-Rieke}\\
  $^2$  From  DSR2012\\
  $*$ In  parenthesis are galaxies in our sample in common with  DSR2012.
 \end{minipage}

\end{table*}

\begin{figure*}
 
  \begin{minipage}{180mm}

    \begin{center}

      \includegraphics[width=29mm]{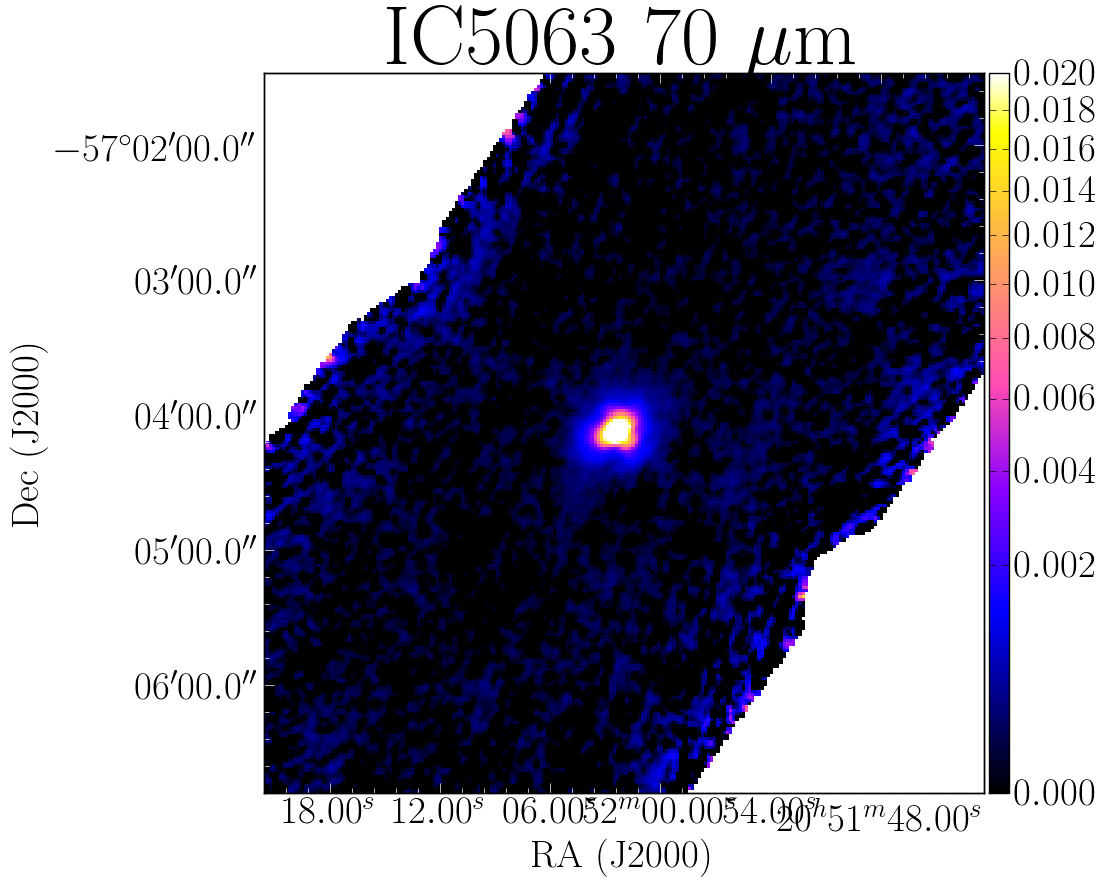}{\hspace{0cm}}
      \includegraphics[width=29mm]{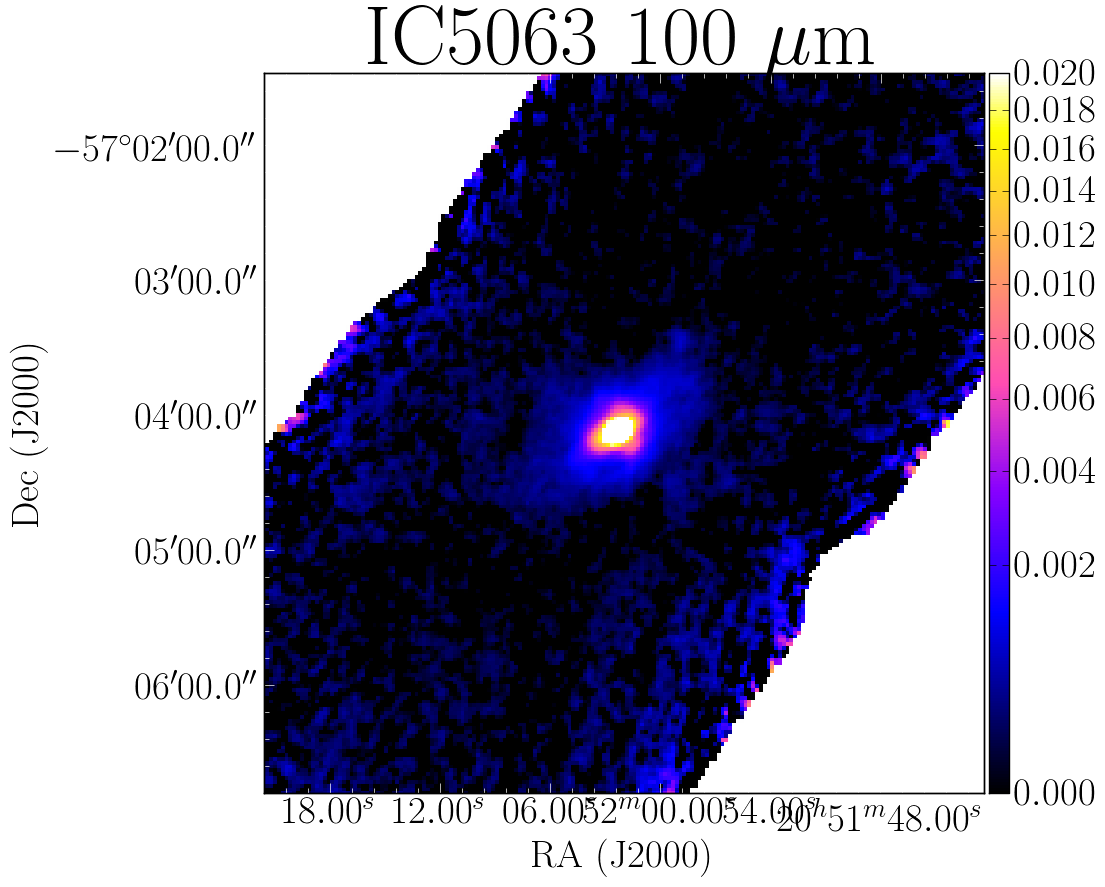}{\vspace{0cm}}
      \includegraphics[width=29mm]{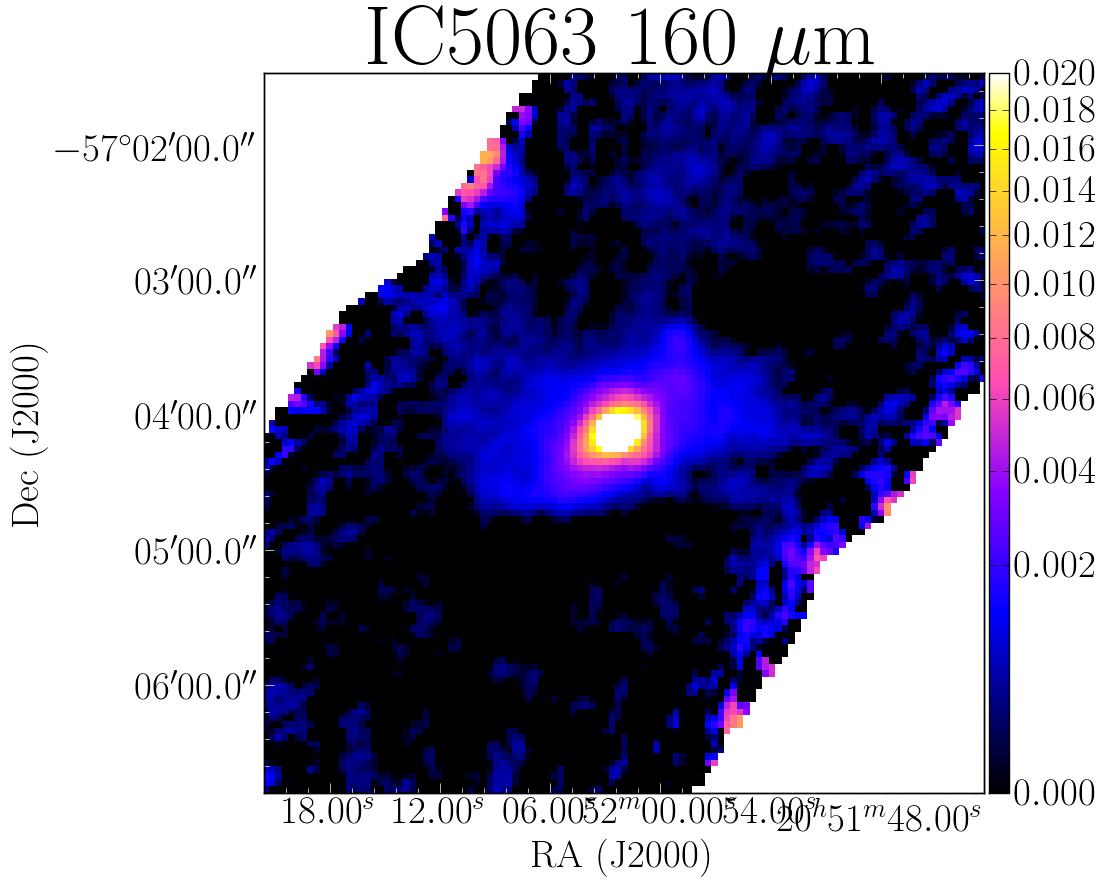}{\vspace{0cm}}
      \includegraphics[width=29mm]{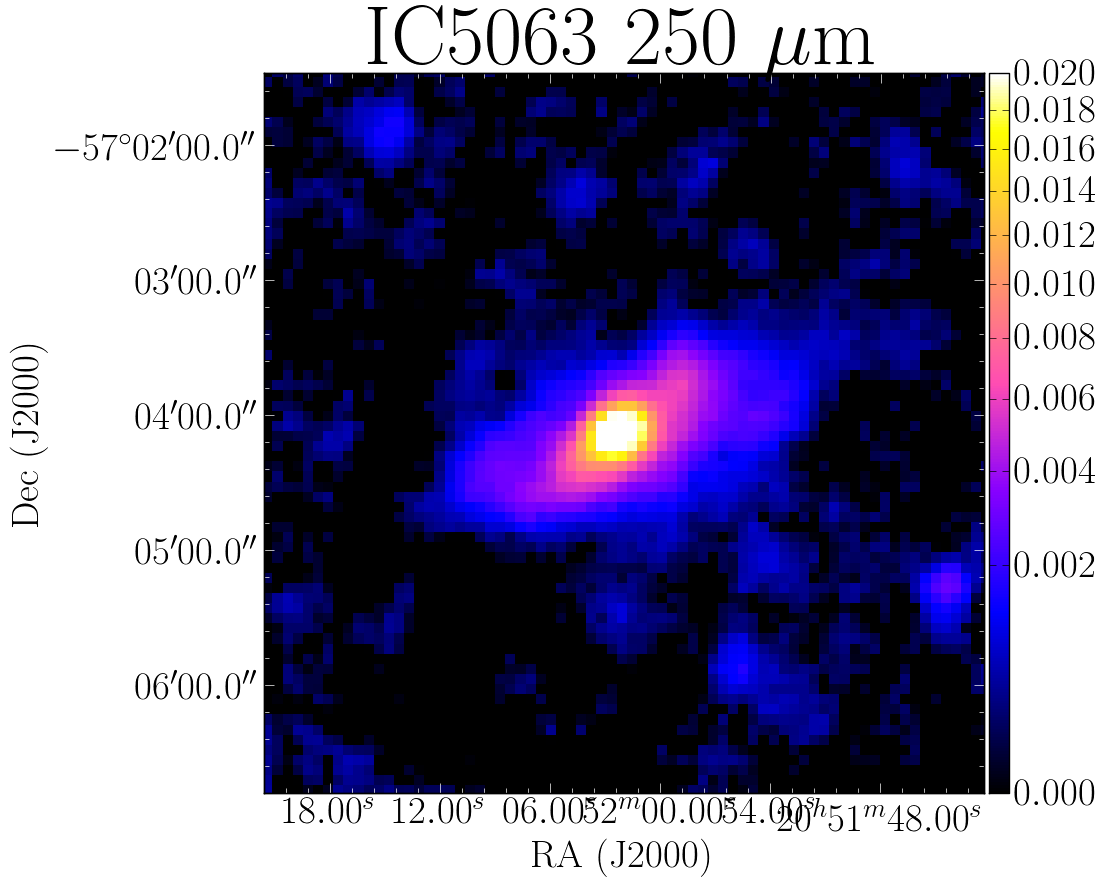}{\vspace{0cm}}
      \includegraphics[width=29mm]{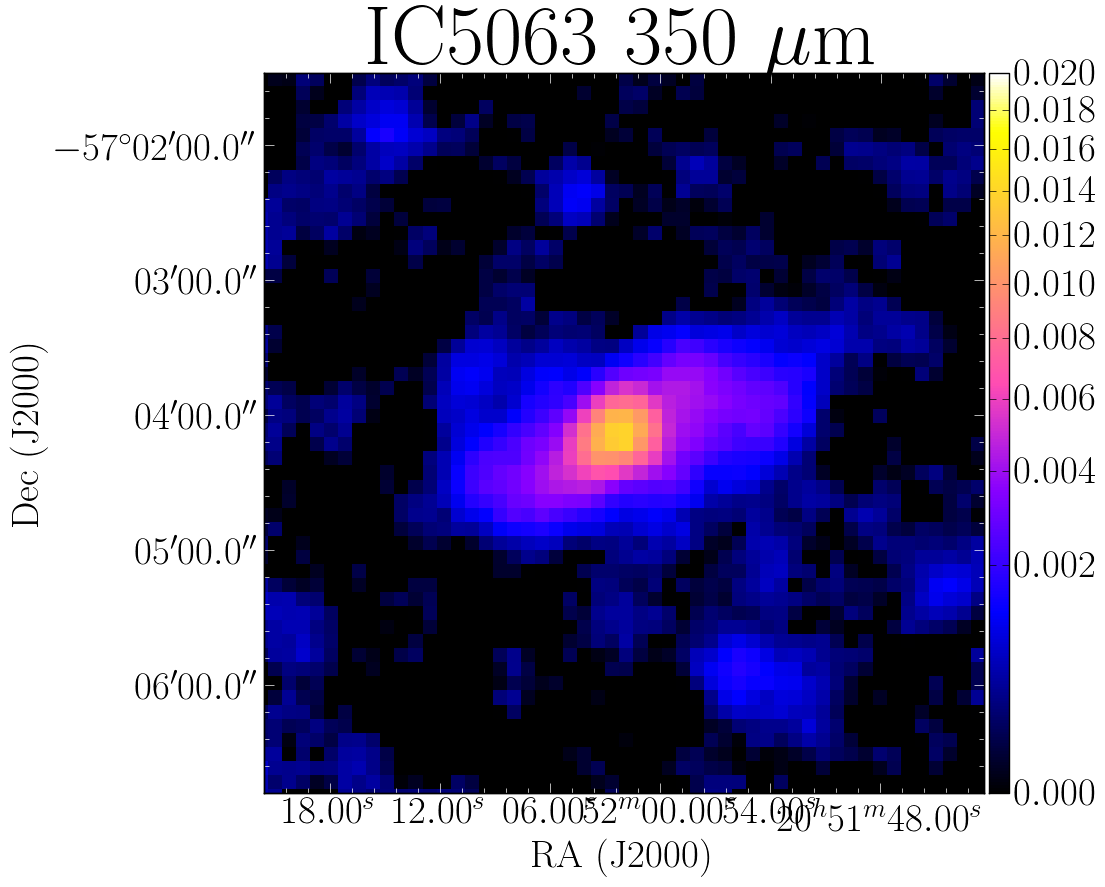}{\vspace{0cm}}
      \includegraphics[width=29mm]{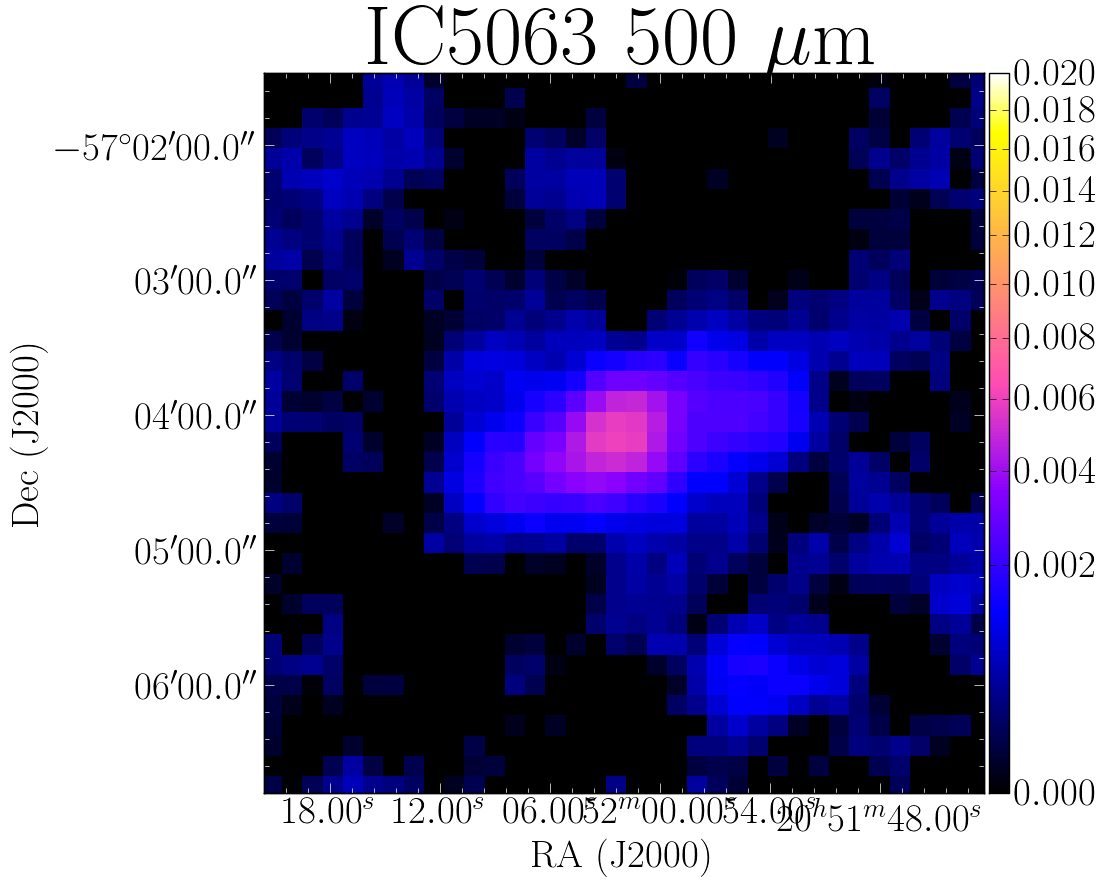}{\vspace{0cm}}

      \caption{Example of the mosaics of IC~5063 in the PACS 70, 100, and $160\,\mu$m bands (the three most
        left panels, left to right), and in the SPIRE 250, 350, and 500 $\mu$m bands (the three
        most right panels from left to right).  The images are shown in a square root scale. The rest of the galaxies are
      available in the online version.}\label{ExampleHerschelimages}

    \end{center}
 
  \end{minipage}

\end{figure*}

We also compared the luminosity distance of our sample those
of the 84 galaxies in  DSR2012. They selected
Seyfert galaxies in the RSA sample with {\it Spitzer}/IRS observations of the $11.3\,\mu$m PAH feature. We used, as 
 DSR2012, the luminosity distance obtained from 
the Nasa Extragalactic Database
(NED\footnote{http://ned.ipac.caltech.edu/}) using the corrected redshift
to the reference frame defined by the Virgo cluster, the Great Attractor and 
the Shapley supercluster. The right top panel of
Fig.~\ref{sample-comparison} shows the luminosity distance 
distributions for the 84 galaxies in  DSR2012 and the 33 galaxies of our sample
 (median luminosity distance of 30 Mpc). 
There are 10 Seyfert galaxies in our sample  not in  
 DSR2012 selection  but  in the RSA catalogue.
Again, from Fig.~\ref{sample-comparison} and a K-S test (p=0.87) our sample
is not statistically significantly different in terms of the luminosity distance.

 Since the main goal of this paper  is to select galaxies with evidence
  of strong contribution of the AGN at 70 $\mu$m emission, in Section~\ref{subsection-SFR}
 we will compare the nuclear SFR from MIR spectroscopy
  and {\it Herschel} 70 $\micron$ photometry.  We
  therefore compared the nuclear SFR values obtained by  DSR2012
for all their sample and the 23 galaxies in common with our sample. 
The  bottom left panel of Fig.~\ref{sample-comparison}
shows the nuclear SFR distribution  in  DSR2012 and our sample with 
 DSR2012 SFR values  (median of 0.32 $M_{\sun} yr^{-1}$). A K-S test  shows that both
samples are  not statistically different in terms of the SFR ($p=0.42$).

We finally  compared the AGN bolometric luminosity distributions
for the galaxies in  DSR2012 and our galaxies with 
their values. This comparison is shown in the right bottom panel 
of Fig.~\ref{sample-comparison}.   Clearly, our sample only
includes the most luminous AGN $\rm L_{bol} > 10^{43}\,{\rm erg\,s}^{-1}$
   when compared to the  DSR2012  RSA sample.
We find a p value for the K-S test of $p=0.03$ when comparing with the galaxies in common and 
$p=0.05$ when comparing with all our sample.  
  This is because in general low-luminosity
  AGN are not bright in the mid-IR and thus few meet our requirement of
  having high angular
  resolution MIR spectroscopy obtained from the ground \citep[see][]{AAH2015}.
   Even in the mid-infrared the low-luminosity AGN do not 
  dominate the emission at least in the 40 per cent of the cases \citep{GonzalezMartin2015}.
Table~\ref{tabla-comparacion-muestras} 
summarizes the statistics of the above comparisons.

\section{Herschel observations}
\label{Section-observations}

In this section we describe the process of obtaining the PACS and 
SPIRE images from the {\it Herschel} archive,  the data reduction 
and the aperture photometry.

The PACS instrument has three bands  referred to as  blue, green, and red 
with  central wavelengths of 70, 100, and 160 $\mu$m, respectively. The nominal
angular resolutions are  5.6, 6.8, and 11.3\,arcsec (FWHM). The red band can be combined with the blue
or the green band, for simultaneous observations. 

The SPIRE instrument has three bands centred at
250, 350, and 500 $\mu$m, with beam sizes of 18.1, 25.2, and 36.9\,arcsec (FWHM),
respectively. 
The three bands are observed simultaneously.

\subsection{Data reduction}

 We compiled all the PACS and SPIRE images for our 33 galaxies using the
Herschel Science Data Archive
(HSA)\footnote{http://www.cosmos.esa.int/web/herschel/science-archive}.
First, we processed the raw level 1 data with the Herschel interactive
pipeline environment software (HIPE) version 13 \citep{Ott} to obtain the
flux calibrated timelines of each bolometer. The standard HIPE pipeline
corrects for instrumental effects and attaches pointing information to
the timelines.
Then, we combined these timelines using Scanamorphos version 24 \citep{Roussel}.
Scanamorphos subtracts low-frequency noise (thermal and
non-thermal), masks high-frequency glitches of the data, and projects
the timelines into a spatial grid.
For the spatial grids, we selected a pixel size of FWHM/4 for each band.
These values provide a good compromise between angular resolution and
sampling (> 10 samples per pixel).
We discarded mosaics at $100\,\mu$m where the 
galaxy reaches the edge of the image only in these band 
because we could not obtain
reliable measurements of the background and could not 
obtain the integrated flux in the same aperture as the other bands.
Only 22 of the 33 galaxies (67 per cent of the sample) have 
observations at $100\,\mu$m after discarding these images.
The list of all the observations used
 can be found in Table~\ref{table-observing-programs}.
Fig.~\ref{ExampleHerschelimages} shows an example of the mosaics of one
  galaxy in our sample observed in all PACS and SPIRE bands. The mosaics for
the rest of the sample are available online. 
 The number
of images used for each band is given in Table~\ref{tabla-FWHM}.

\begin{table*}
 \caption{{\it Herschel}/PACS and SPIRE observing programs. The full table is
 available in the online version.}
 \label{table-observing-programs}
 \begin{tabular}{@{}ccccc}
  \hline
  Galaxy & Observation ID & Instrument & Wavelengths ($\mu$m) & PI \\
  \hline
  ESO~323-G077&1342236922&PACS&70, 160&Mushotzky R.\\
  ESO~323-G077&1342236923&PACS&70, 160&Mushotzky R.\\
  ESO~323-G077&1342236202&SPIRE&250, 350, 500&Mushotzky R.\\
  IC~5063&1342216469&PACS&70, 160&S\'anchez-Portal M.\\
  IC~5063&1342216470&PACS&70, 160&S\'anchez-Portal M.\\
  IC~5063&1342216471&PACS&100, 160&S\'anchez-Portal M.\\
  IC~5063&1342216472&PACS&100, 160&S\'anchez-Portal M.\\
  IC~5063&1342206208&SPIRE& 250, 350, 500&S\'anchez-Portal M.\\
  
  \hline
  
  \end{tabular}

\end{table*}

\subsection{Aperture photometry}

Circular aperture photometry of the galaxies was carried out using  HIPE version 13. 
For every galaxy and every band we performed the photometry 
for radii of 1\,kpc, and 2\,kpc, as well as for the total galaxy. To determine 
the size in arcsec corresponding to 1\,kpc and 2\,kpc we took into account the 
luminosity distance of each galaxy. The aperture corresponding to the 
total galaxy was determined by visual inspection of the radial
profiles in all the bands. In Table~\ref{tabla-fotometria-total}
we list the apertures used for the total photometry.
We imposed that the aperture had to be the same for all
the bands. In some galaxies, part 
of the galaxy seen in the SPIRE bands is outside the aperture used 
for the photometry. To avoid errors in the photometry due to
the small size of the aperture, we only consider fluxes from nuclear apertures 
with a diameter higher than 1.5 times the angular resolution
of each band. The photometric 
error of each image is determined by
placing six apertures on the background around the source and
 measuring the  standard deviation  per pixel of the background within these apertures.
  The error in the flux was then computed by
  multiplying the  standard deviation  per pixel by the number of  pixels of the aperture. 

\begin{table*}
 \caption{ FWHMs measured in the six FIR bands.}
 \label{tabla-FWHM}
 \begin{tabular}{lccccccccccc}
  \hline
  &\multicolumn{6}{c}{FWHM ($\arcsec$)}&\multicolumn{3}{c}{Number of images}& FWHM (kpc)\\
  Galaxy&70 $\mu$m & 100 $\mu$m & 160 $\mu$m 
  & 250 $\mu$m &350 $\mu$m & 500 $\mu$m
  &70 $\mu$m & 100 $\mu$m & 160 $\mu$m&70 $\mu$m  \\
  \hline
  ESO~323-G077 & 6.7 &   & 12.4 & 19.2 & 25.4 & 36.7 & 2 & - & 2 &1.9 \\
  IC~5063 & 6.0 & 7.8 & 13.1 & 22.6 & 33.6 & 51.9 & 2 & 2 & 4 &1.4\\
  Mrk~1066 & 5.9 & 7.2 & 11.8 & 19.5 & 25.9 & 38.4 & 2 & 2 & 4 & 1.4\\
  NGC~1068 & 6.6 &   & 21.3 & 36.1 & 39.9 & 48.0 & 2 & - & 2 & 0.5 \\
  NGC~1320 & 6.1 & 8.4 & 14.7 & 22.2 & 28.5 & 41.2 & 2 & 2 & 4 & 1.0 \\
  NGC~1365 & 10.3 & 11.9 & 17.4 & 23.7 & 29.9 & 39.9 & 2 & 2 & 4 & 1.1\\
  NGC~1386 & 6.1 & 8.0 & 14.5 & 23.8 & 30.4 & 39.8 & 2 & 2 & 4 & 0.3\\
  NGC~1808 & 9.4 & 11.2 & 15.3 & 21.3 & 28.3 & 39.8 & 2 & 2 & 4 & 0.6\\
  NGC~2110 & 6.6 & 8.3 & 13.2 & 20.6 & 27.2 & 46.0 & 2 & 2 & 4 &1.0\\
  NGC~2273 & 6.1 & 7.6 & 12.2 & 20.4 & 27.9 & 40.5 & 2 & 2 & 4 &0.8\\
  NGC~2992 & 7.9 & 9.5 & 14.2 & 21.1 & 28.6 & 39.9 & 2 & 2 & 4 &1.3 \\
  NGC~3081 & 6.2 & 8.1 & 13.3 & 22.9 & 37.6 & 65.1 & 2 & 2 & 4 &1.0\\
  NGC~3227 & 6.3 &  & 12.7 & 22.4 & 31.9 & 46.4 & 2 & - & 4 & 0.6\\
  NGC~3281 & 5.8 & 7.3 & 11.9 & 19.5 & 26.8 & 39.4 & 2 & 2 & 4 &1.3 \\
  NGC~3783 & 5.7 & 7.4 & 22.0 & 52.7 & 57.8 & 60.8 & 2 & 2 & 4 &1.0\\
  NGC~4051 & 6.0 &   & 14.7 & 31.2 & 49.3 & 183.6 & 2 & - & 4 & 0.4\\
  NGC~4151 & 5.9 &   & 13.7 & 31.3 & 44.5 & 141.7 & 2 & - & 4 &0.6\\
  NGC~4253 & 5.8 & 7.2 & 11.9 & 19.7 & 28.3 & 66.0 & 2 & 2 & 4 &1.7\\
  NGC~4258 & 49.9 & 71.4 & 71.3 & 90.8 & 73.3 & 97.7 & 3 & 1 & 4 &1.9\\
  NGC~4388 & 6.9 &   & 15.7 & 26.6 & 36.4 & 51.8 & 2 & - & 2 &0.6\\
  NGC~4507 & 5.7 &   & 12.8 & 28.5 & 44.2 & 58.1 & 2 & - & 2 &1.6\\
  NGC~4579 & 7.0 & 9.3 & 110.6 & 27.4 & 39.6 & 138.5 & 2 & 2 & 4 &0.6\\
  NGC~4594 & 6.9 & 11.8 & 69.7 & 57.9 & 43.8 & 47.0 & 2 & 2 & 4 &0.4\\
  NGC~4725 & 10.7 & 24.1 & 22.6 & 73.0 & 87.6 & 308.8 & 2 & 2 & 4 &1.4\\
  NGC~5135 & 6.8 & 8.1 & 12.3 & 19.7 & 26.3 & 40.6 & 2 & 2 & 4 &1.9\\
  NGC~5347 & 5.6 & 7.4 & 12.7 & 25.6 & 43.9 & 66.3 & 2 & 2 & 4 &1.1\\
  NGC~5506 & 6.0 & 7.4 & 12.0 & 20.2 & 27.4 & 42.8 & 2 & 2 & 4 &0.9\\
  NGC~7130 & 5.8 &   & 12.9 & 20.8 & 27.6 & 39.7 & 2 & - & 2 &1.9\\
  NGC~7172 & 7.1 &   & 13.4 & 20.8 & 27.3 & 41.2 & 2 & - & 2 &1.3\\
  NGC~7213 & 6.5 &   & 20.1 & 53.0 & 61.4 & 60.3 & 2 & - & 2 &0.8\\
  NGC~7465 & 7.3 & 9.2 & 14.1 & 22.1 & 30.0 & 47.2 & 2 & 4 & 6 &1.0 \\
  NGC~7479 & 5.6 & 7.3 & 12.3 & 21.6 & 32.3 & 47.9 & 2 & 2 & 4 &0.9\\
  NGC~7582 & 6.6 &   & 12.3 & 19.8 & 26.7 & 42.6 & 2 & - & 4 &0.7\\
  
  \hline
  
  \end{tabular} 
  
  Notes.--- All the galaxies have one 
  image at the SPIRE bands  except NGC~4258 and 
  NGC~7465, which have two images.

  \end{table*}

We measured the FWHM for each galaxy at each band using  Moffat with the Image
Reduction and Analysis Facility ({\sevensize IRAF}\footnote{IRAF is
  distributed by the National Optical Astronomy Observatory, which is
  operated by the Association of Universities for Research in
  Astronomy (AURA), Ibc., under cooperative agreement with the
  National Science Foundation.}).  We list in Table~\ref{tabla-FWHM}
  the measured FWHM in all six {\it Herschel} bands. We then applied an aperture 
  correction using HIPE to the fluxes of those galaxies 
 that are  point sources and quasi point sources. 
 We did not apply an aperture correction to the extended sources.
We consider  that a source is quasi point-like at one particular band (although
it may have extended emission) when the measured 
FWHM (see Table~\ref{tabla-FWHM}) is less than the FWHM of the instrument 
plus the number of arcseconds in one pixel. 
For a typical galaxy in our sample at  30\,Mpc, the aperture correction at 70 $\mu$m is
1.48  and 1.22 for  photometry radio of 1 and 2\,kpc, respectively.
 For sources that are quasi point-like the
  unresolved fluxes are probably slightly overestimated, as we shall see in
Section~\ref{Unresolvedemission}. 
This is due to the instrumental  Point Spread Function (PSF) that causes  that a fraction of the
external emission of the aperture  to be observed in the inner region.
 However, although this correction cannot be calculated, 
it should be within the flux error and it does not modify our results.

To obtain the final error of the fluxes we
added in quadrature the photometric calibration uncertainties (10 per cent) to the error calculated above.
In Tables~\ref{tabla-fotometria-1kpc}, \ref{tabla-fotometria-2kpc} and \ref{tabla-fotometria-total} we provide
the fluxes for all galaxies and all bands for $r=1\,$kpc, $r=2\,$kpc, and the total galaxy.
We also list, for every galaxy, the size in arcsec for the aperture used. As
can be seen from this table, the photometric calibration uncertainty dominates in most cases.

\begin{table*}
 \caption{Aperture photometry for r = 1 kpc. The full table is
 available in the online version.}
\label{tabla-fotometria-1kpc}
 \begin{tabular}{@{}cccccccc}
  \hline
   Galaxy & radius & Flux  &   Flux &  Flux &  Flux &  Flux  & Flux\\
  & (arcsec) & 70$\mu$m (Jy)& 100$\mu$m (Jy) & 160$\mu$m (Jy)& 250$\mu$m (Jy)& 350$\mu$m (Jy)& 500$\mu$m (Jy)\\
  \hline

  ESO~323-G077 & 3.43 & &    &    &    &    &   \\
  IC~5063  & 4.13 &  &  &    &    &    &   \\
  Mrk~1066 & 4.21 & 12$\pm$1 & &    &    &    &   \\
  NGC~1068 & 14.32 & 164$\pm$16 &    & 88$\pm$9 & 28$\pm$3 & & \\
  NGC~1320 & 5.81 & 1.8$\pm$0.2 & 2.0$\pm$0.2&    &    &    &   \\

  \hline
 \end{tabular}

\end{table*}

\begin{table*}
 \caption{Aperture photometry for r = 2 kpc. The full table is
 available in the online version.}
\label{tabla-fotometria-2kpc}
 \begin{tabular}{@{}cccccccc}
  \hline
  Galaxy  & radius  & Flux   &   Flux  &  Flux  &  Flux  &  Flux   &  Flux\\
   & (arcsec)  & 70$\mu$m (Jy) & 100$\mu$m (Jy)  & 160$\mu$m (Jy) & 250$\mu$m (Jy) & 350$\mu$m (Jy) & 500$\mu$m (Jy)\\
  \hline

  ESO~323-G077   & 6.85  & 6.9$\pm$0.7  &     & & &  & \\
  IC~5063   & 8.27  & 4.5$\pm$0.5  & 3.6$\pm$0.4 && &  & \\
  Mrk~1066  & 8.42  & 12$\pm$1  & 12$\pm$1 & &  &  & \\
  NGC~1068  & 28.65  & 253$\pm$25  &   & 186$\pm$19  & 65$\pm$7  & 23$\pm$2  & 8.0$\pm$0.8\\
  NGC~1320  & 11.62  & 2.2$\pm$0.2  & 2.8$\pm$0.3  & 2.3$\pm$0.2 & &  & \\

  \hline
 \end{tabular}

\end{table*}

\begin{table*}
 \caption{Integrated photometry. The full table is
 available in the online version.}
 \label{tabla-fotometria-total}
 \begin{tabular}{@{}cccccccc}
  \hline
  Galaxy  & radius  & Flux   &   Flux  &  Flux  &  Flux  &  Flux   &  Flux\\
   & (arcsec)  & 70$\mu$m (Jy) & 100$\mu$m (Jy)  & 160$\mu$m (Jy) & 250$\mu$m (Jy) & 350$\mu$m (Jy) & 500$\mu$m (Jy)\\
  \hline

  ESO~323-G077&50 & 6.8$\pm$0.8  &                & 7.1$\pm$0.8 & 3.2$\pm$0.3 & 1.3$\pm$0.1 & 0.43$\pm$0.04 \\
  IC~5063   & 60  & 4.7$\pm$0.5  & 4.6$\pm$0.5  & 3.7$\pm$0.4 & 2.1$\pm$0.2 & 0.9$\pm$0.1 & 0.30$\pm$0.04 \\
  Mrk~1066  & 50  & 12$\pm$1 & 13$\pm$1 & 8.5$\pm$0.9 & 3.0$\pm$0.3 & 1.1$\pm$0.1 & 0.33$\pm$0.04 \\
  NGC~1068  & 230 & 292$\pm$30  &                   & 299$\pm$30 & 117$\pm$12 & 45$\pm$5 & 15$\pm$2 \\
  NGC~1320  & 50  & 2.4$\pm$0.3  & 3.4$\pm$0.4  & 3.0$\pm$0.4 & 1.5$\pm$0.2 & 0.58$\pm$0.07 & 0.20$\pm$ 0.03\\

  \hline
 \end{tabular}

\end{table*}

\begin{figure}
  \includegraphics[width=0.45\textwidth]{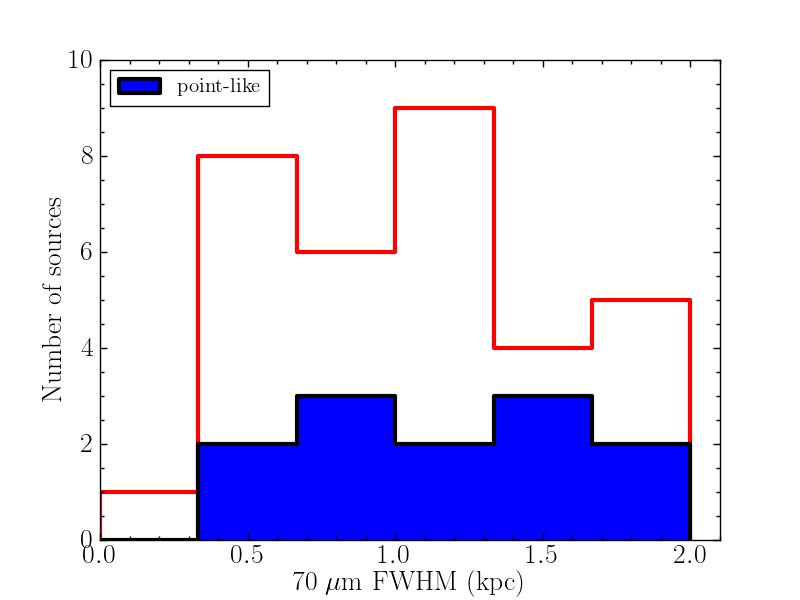}
  \caption{Distribution of the measured FWHM at $70\,\mu$m in kpc for our sample (open
    histogram). The filled histograms are those nuclei which appear unresolved
    at this wavelength, that is, with FWHM$<$6\,arcsec.} 
  \label{FWHM_70micron_kpc}
\end{figure}

\begin{figure}
  \includegraphics[width=0.45\textwidth]{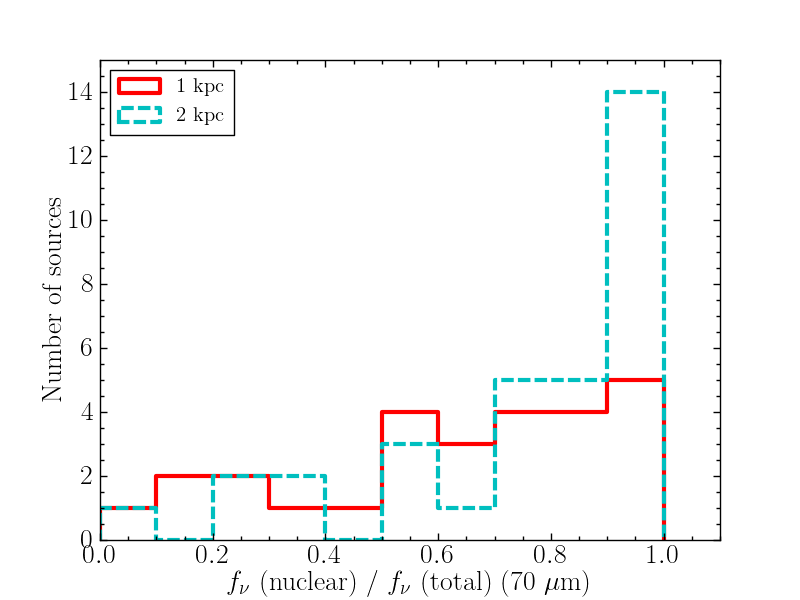}
  \caption{Distribution of the ratio between nuclear  and total flux at 70 $\mu$m
    for $r=1\,$kpc for (27 galaxies, in red) and $r=2\,$kpc 
  (33 galaxies, in dashed cyan).} 
  \label{flujo-nuclear-total-blue}
\end{figure}

\section{Results}
\label{Section-results}

In this Section we analyse the FIR properties of our sample, such as, 
the unresolved $70\,\mu$m emission, the FIR colour distributions, 
the results from fitting the data to a grey body and the 
SFR nuclear and extranuclear obtained from the 70 $\mu$m data.

\subsection{Unresolved $70\,\mu$m emission}\label{Unresolvedemission}

Of the 33 galaxies in our sample,  12 nuclei (31 per cent) appear point-like 
at 70 $\mu$m, i.e., have FWHM $<$ 6  arcsec (see Table~\ref{tabla-FWHM}). 
  As can be seen from Fig.~\ref{FWHM_70micron_kpc}, the $70\,\mu$m emission for
  those nuclei seen as point sources originates in
  regions with sizes (diameters) of less than $\sim 0.5-2\,$kpc, with a median size  of 1.3\,kpc. These
  values are, however, comparable to those nuclei in our sample that appear clearly extended
  at $70\,\mu$m  (median size of 1.0\,kpc).
The fraction of RSA Seyfert galaxies with unresolved emission is similar to
that of the {\it Swift}/BAT  hard X-ray selected AGN sample of 
\citet{Mushotzky}. They found that $>$35 per cent of their sources are point-like at $70\,\mu$m  
with typical sizes at $70\,\mu$m of 2\,kpc.
      However, the {\it Swift}/BAT AGN sample is on average at $z\sim 0.025$, compared to the
      average $z=0.007$ of our sample. Therefore, it is likely that the difference in
redshifts between our sample and the {\it Swift}/BAT sample explains the different physical sizes
for the 70$\micron$  emitting nuclear region.

To study  further the unresolved
70 $\mu$m emission we estimated
the contribution of the nuclear region to the total flux at $70\,\mu$m, as 
is shown in Fig.~\ref{flujo-nuclear-total-blue}.
 The median values of the nuclear contributions
to the total flux are  0.68 and 0.86 for $r=1\,$kpc and $r=2\,$kpc, respectively
(see Table \ref{estadistica-contribucion-nuclear}).

Of the 27
galaxies with  measurements of the nuclear flux in $r=1\,$kpc at 70 $\mu$m,   20 (74 per cent) have 
a nuclear 1 kpc contribution to the total flux greater than 50 per cent. Of the 
33 galaxies with nuclear fluxes within $r=2\,$kpc at 70 $\mu$m,  28 (85 per cent) have 
a nuclear $r=2\,$kpc contribution greater than 50 per cent.
The values obtained for 1\,kpc and 2\,kpc are  similar because 
88 per cent of the galaxies required an aperture correction for the fluxes.
Our results are in agreement with \citet{Mushotzky}.
They found that 92.5 per cent (274 out of 296 galaxies) of their sample had a point source contribution greater
than 50 per cent of the total flux at 70 $\mu$m. They found a slightly higher percentage
because they performed the photometry with an aperture of 6  arcsec for the PACS 
70 $\mu$m images for all the galaxies, independently of their distance. 
 As their galaxies are more distant than our sample, then the regions 
  for their nuclear photometry are larger than our $r=1\,$kpc and $r=2\,$kpc nuclear
regions. We note that \citet{Mushotzky} also performed an aperture correction 
to the fluxes.

\begin{table}
 \caption{Contribution of 
the nuclear regions of r = 1\,kpc and 2\,kpc to the total flux at 70 $\mu$m in RSA Seyferts}
 \label{estadistica-contribucion-nuclear}
 \begin{tabular}{lcccc}
  \hline
  Quantity &Number & Mean& $\sigma$ & Median\\
  \hline
  $f_\nu$(r=1kpc)/$f_\nu$(total) & 27 & 0.63  & 0.28 & 0.68 \\
  $f_\nu$(r=2kpc)/$f_\nu$(total) & 33 &  0.76   &  0.26  &  0.86 \\
  \hline
 \end{tabular}

\end{table}

\subsection{FIR colours}

\begin{figure}
  \includegraphics[width=0.45\textwidth]{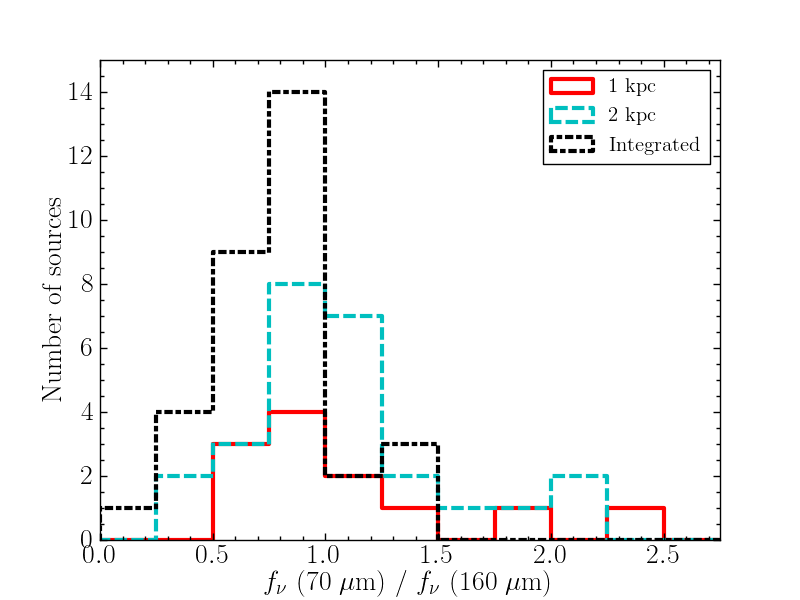}
  \caption{Distribution of  the  $f_\nu$(70 $\mu$m)/$f_\nu$(160 $\mu$m) flux ratios
  within $r=1\,$kpc (12 galaxies, solid red line), within $r=2\,$kpc (26 galaxies,
  dashed cyan line), and for the integrated galaxy (33 galaxies, dash-dot black line)
  for the RSA Seyferts.}
  \label{distribucion-color}
\end{figure}

\begin{table}
  \caption{Comparison of the observed $f_\nu$(70 $\mu$m)/$f_\nu$(160 $\mu$m) flux ratios
  for different samples}
 \label{estadistica-colores}
 \begin{tabular}{lcccc}
  \hline
  Region &Number& Mean& $\sigma$ & Median\\
  \hline
  \multicolumn{5}{c}{RSA Seyferts}\\
  \hline
  $r=1\,$kpc &  12 &  1.13 &  0.52 &  0.98 \\
  $r=2\,$kpc & 26 &  1.08 & 0.44 & 1.01 \\
  Integrated      & 33 &  0.78 & 0.30 &  0.80 \\
    \hline
  \multicolumn{5}{c}{{\it Swift}/BAT AGN}\\
  \hline
  Integrated &  258 &  0.79 &  0.44 &  0.68\\
  \hline
  \multicolumn{5}{c}{PG quasars}\\
  \hline
  Integrated &  68 &  1.41 &  0.87 &  1.25\\
  \hline
  \multicolumn{5}{c}{KINGFISH galaxies}\\
  \hline
  Integrated normal &  29 &  0.75 &  0.45 &  0.66\\
  Integrated AGN &  31 &  0.51 &  0.27 &  0.43\\
  Integrated all &  60 &  0.63 &  0.39 &  0.50\\
  \hline
 \end{tabular}

\end{table}

 We first start by discussing the $f_\nu$(70 $\mu$m)/$f_\nu$(160 $\mu$m) flux ratio 
  since we  can still obtain nuclear values for a significant fraction of galaxies in our sample.
   It provides information about the peak of the galaxy spectral
  energy distribution (SED) and is a proxy for the dust temperature \citep{Skibba2011,Melendez}. 
Figure~\ref{distribucion-color} shows the distributions of this ratio for 
$r=1\,$kpc and $r=2\,$kpc, and for the entire galaxy. As can be seen
  from this figure (see Table~\ref{estadistica-colores} for the statistical information),
the flux ratios for the nuclear regions  tend to be higher than
  those measured for the entire galaxies. This is in good agreement with the 
decreasing $f_\nu$(70 $\mu$m)/$f_\nu$(160 $\mu$m) flux ratios
with galactocentric radius found for M81 and M83 
\citep{Bendo}.  These authors suggested that this ratio tends to be more strongly
  influenced by star forming regions than other FIR ratios involving longer wavelengths.
  Therefore, the higher nuclear ratios in our sample could be due to higher star formation activity
  but also to higher  dust temperature due to AGN heating (see further discussion
  in Section~\ref{Section-criteria}).

 For the {\it Swift}/BAT AGN sample, \citet{Melendez} measured 
a mean integrated $f_\nu$(70 $\mu$m)/$f_\nu$(160 $\mu$m) flux ratio of $0.81\pm0.43$
for Seyfert 1s and $0.77\pm0.47$ for Seyfert 2s. These are fundamentally
the same as for our sample of optically
selected AGN. 
\citet{Melendez} also compared their observed colours with  predictions from three
different torus models and found that the torus models cannot produce 
$f_\nu$(70 $\mu$m)/$f_\nu$(160 $\mu$m)  ratios of less than unity.
This suggests that  the 70 $\mu$m and 
160 $\mu$m integrated emission is dominated by the host galaxy.

\citet{Melendez} also compared the {\it Swift}/BAT AGN FIR colour distribution
with the Key Insights on Nearby Galaxies: a 
Far-Infrared Survey with {\it Herschel} (KINGFISH) sample of nearby galaxies
(\citealt{Dale}). For this comparison, they only selected the normal galaxies
in the KINGFISH sample using the spectral classification of \cite{Moustakas2010},
and demonstrated that the BAT AGN FIR colours are statistically indistinguishable from
those of normal galaxies. As our colour distribution for the total galaxies is compatible 
with the BAT AGN colour distribution, the colours of our galaxies are 
 not statistically different from those of normal  galaxies   ($p=0.20$) comparing our 
galaxies with   normal galaxies in the KINGFISH 
sample (see also Table~\ref{estadistica-colores}).

\cite{Shimizu2015} also studied the BAT sample and found anomalous colours for 6 BAT
AGN with $f_\nu$(250 $\mu$m)/$f_\nu$(350 $\mu$m) < 1.5 and 
$f_\nu$(350 $\mu$m)/$f_\nu$(500 $\mu$m) < 1.5. They suggested that this might
be an excess based on the synchrotron emission from the jet
or the corona from the accretion disks. We do not find these 
anomalous colours for our galaxies.

\begin{figure*} 
  \begin{center}

      \includegraphics[width=0.49\textwidth]{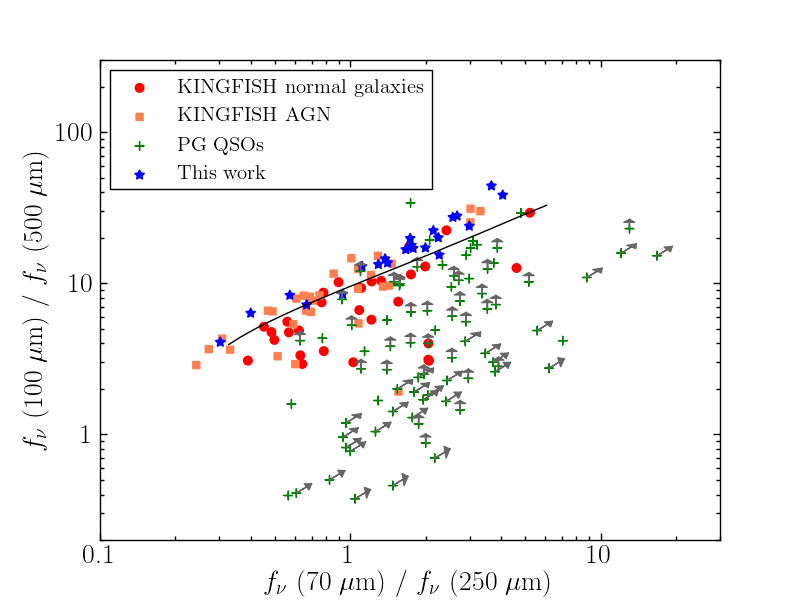}{\vspace{0cm}}
      \includegraphics[width=0.49\textwidth]{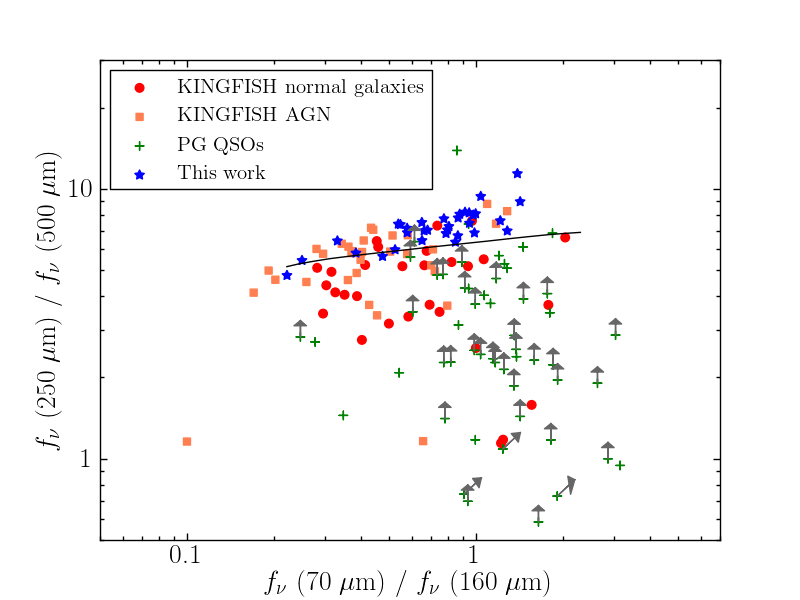}{\vspace{0cm}}

      \caption{Colour-colour diagrams for our RSA Seyfert galaxies (blue stars), the  normal galaxies
      from the KINGFISH sample
        (red circles) of \citet{Dale}, the AGN galaxies from the KINGFISH sample
        (coral circles)  and the PG quasar sample (green crosses) of \citet{Petric}.
        The grey arrows are quasars with an upper limit to the flux. The solid lines indicate
        the models from \citet{Dale-Helou}. These 
  models of SEDs are derived from the average global trends for a sample
  of normal star-forming galaxies.}

  \label{figuras-tipo-dale}

  \end{center}
 
\end{figure*}

We also compare the  other integrated FIR colours of our galaxies with the KINGFISH 
sample of \citet{Dale}  (including normal galaxies and AGN)
and with the 85 nearby (z$\leqslant$0.5) quasars
from the Palomar-Green (PG) sample of \citet{Petric}. In 
Fig.~\ref{figuras-tipo-dale}  we show the
$f_\nu$(100 $\mu$m)/$f_\nu$(500 $\mu$m) vs. $f_\nu$(70 $\mu$m)/$f_\nu$(250 $\mu$m) colour-color
diagram (left)
and $f_\nu$(250 $\mu$m)/$f_\nu$(500 $\mu$m) vs. $f_\nu$(70 $\mu$m)/$f_\nu$(160 $\mu$m)    
colour-colour diagram (right) for our galaxies,
the KINGFISH sample and the PG quasar sample. Most of the PG quasars
only have upper limits of the fluxes for some bands. There are 65
quasars with upper limits at 500 $\mu$m, 29 at 250 $\mu$m, 16 at 160 $\mu$m, 
4 at 100 $\mu$m and 2 at 70 $\mu$m.  We do not display 
quasars with upper limits in the two fluxes involved in a ratio.
In the left panel of Fig.~\ref{figuras-tipo-dale} we show 81 
quasars of the PG sample and in the right panel we show 54 quasars.
We indicate with grey arrows the quasars with upper limits.

 From Fig.~\ref{figuras-tipo-dale}, we observe that the
$f_\nu$(250 $\mu$m)/$f_\nu$(500 $\mu$m) and $f_\nu$(70 $\mu$m)/$f_\nu$(160 $\mu$m)    
ratios of the  RSA Seyfert galaxies
($7.26\pm1.21$ and $0.78\pm0.$30, respectively) 
are, on average, higher than those of the KINGFISH sample
($4.96\pm1.73$ and $0.63\pm0.39$, respectively).
A K-S test indicates that the distributions of the two flux
ratios are statistically different
for our sample and the KINGFISH sample. We obtain  p-values
of $p=0.002$ and $p=10^{-7}$ for the $f_\nu$(70 $\mu$m)/$f_\nu$(160 $\mu$m)    
and $f_\nu$(250 $\mu$m)/$f_\nu$(500 $\mu$m) ratios,  which indicates
that they are statistically drawn from different parent samples.  
This may be due to the large range  of
  morphologies and metallicities of the KINGFISH galaxies, whereas the RSA Seyfert galaxies
 are mostly early type (see Table~\ref{galaxy-sample}). 
The differences in the integrated $f_\nu$(70 $\mu$m)/$f_\nu$(160 $\mu$m) flux ratios 
are not in conflict with the comparison between the BAT AGN sample and the KINGFISH 
sample. \citet{Melendez} compared their galaxies with the normal galaxies in the 
KINGFISH sample. A K-S test indicates that $f_\nu$(70 $\mu$m)/$f_\nu$(160 $\mu$m) ratios 
of \citet{Melendez} and all the KINGFISH sample
are statistically different ($p=0.003$).

\subsection{Grey-body fitting}\label{sec:grey body-fits}

 In this section we estimate the dust temperature
by fitting the {\it Herschel} SED to a grey body.
We use \textsc{sherpa} (\citealt{Doe}), that is,  the 
Chandra Interactive Analysis of Observations ({\it CIAO's}) {\it modelling and fitting
package}  (\citealt{Freeman})   module for  \textsc{python}.
We fit $\nu F_\nu$ versus the rest-frame wavelength,
using the following expression:

\begin{equation}
\nu F_{\nu_{model}} ={A \over {\lambda^{3+\beta}(e^{{h c} \over {\lambda k T}}-1})}{c \over \lambda}   
\end{equation}
where $\lambda$ is the rest-frame wavelength, and the free parameters of the fitting are
the amplitude ($A$), the dust temperature ($T$), and the  dust
  emissivity index ($\beta$). To estimate the 
best fit to the data we first minimized the usual $\chi^2$
statistics leaving all the parameters to vary freely:

\begin{equation}
  \chi^2=\sum_{i}^N{(\nu F_{\nu_i}-\nu F_{\nu_{i,model}})^2\over{\sigma_i^2}}
  \label{ec-Xcuadrado}
\end{equation}
where $N$ is the number of data points, $F_{\nu_i}$ is
the flux at the $i$th  wavelength, $\sigma_i$ is
the uncertainty of the observed flux at the $i$th wavelength, and 
$F_{\nu_{i,model}}$ is the predicted $F_{\nu}$ for 
the $i$th wavelength.  We perform the fits for
  the integrated SEDs as well as for the nuclear $r=1\,$kpc and
  $r=2\,$kpc SEDs. We require four or more data points 
  for the fittings.

\begin{figure*}
    \begin{center}
 
      \includegraphics[width=55mm]{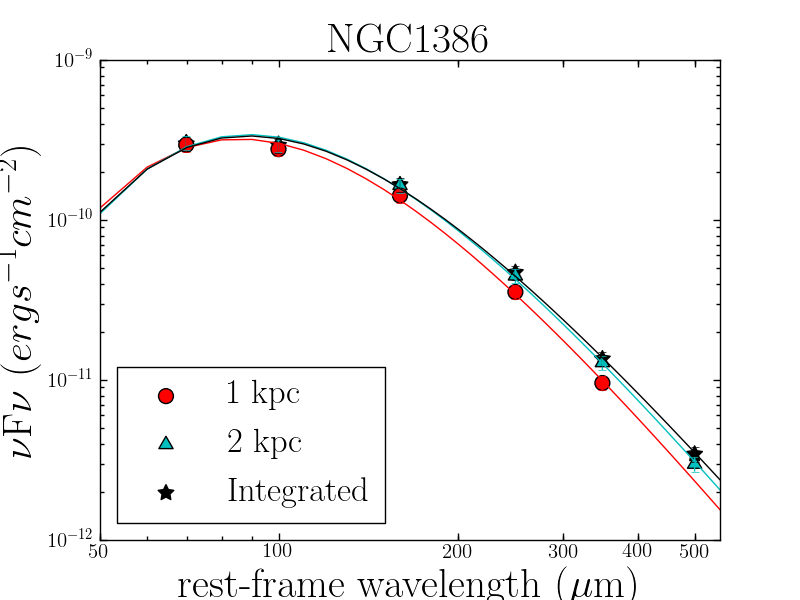}{\vspace{0cm}}
      \includegraphics[width=55mm]{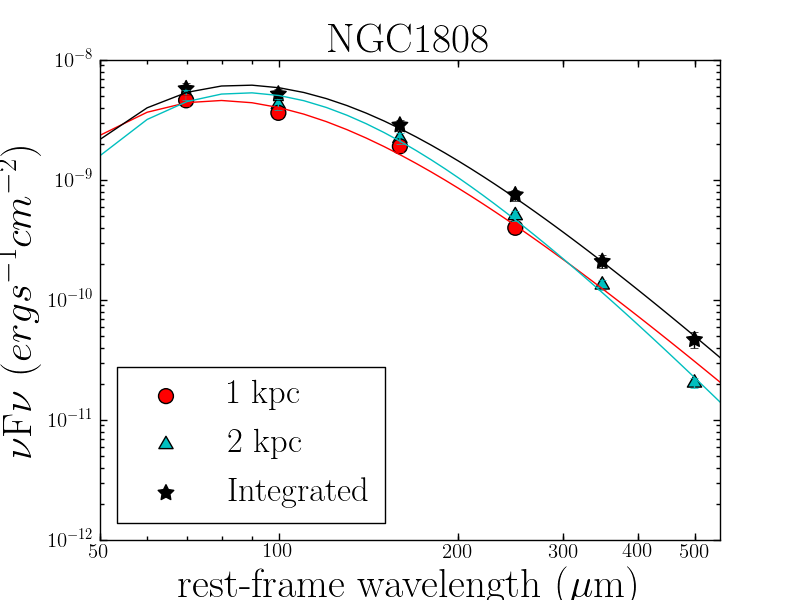}{\vspace{0cm}}
      \includegraphics[width=55mm]{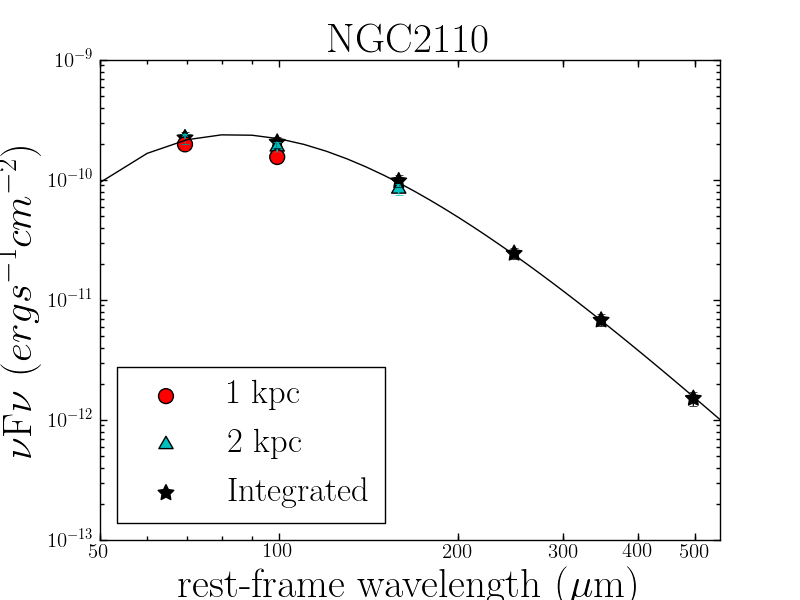}{\vspace{0cm}}

      \caption{Examples of best fits to the FIR SEDs of the galaxies with a grey body with all the
        parameters left free to vary. The red circles are the fluxes for 
        $r=1\,$kpc, the cyan triangles for $r=2\,$kpc and  black stars for the total galaxy.
        The lines indicate the 
        best fit in each region (only if there are 4 or more
        data points), with the colours as for the data points.
The best fits for the rest of the galaxies can be found in the online version.}

  \label{exampleSEDs}

    \end{center}

\end{figure*}

\begin{figure}
  \includegraphics[width=0.45\textwidth]{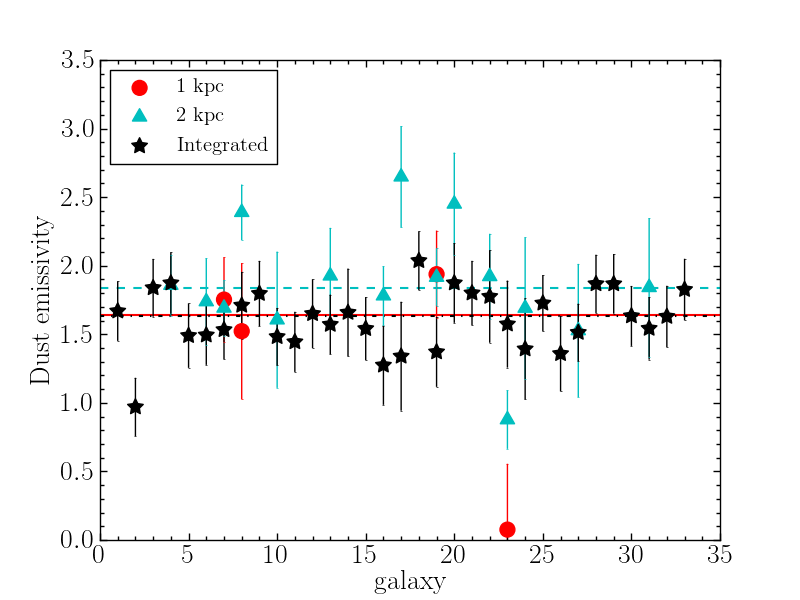}
  \caption{Distribution of  the fitted dust emissivity
      indices $\beta$. 
    For each galaxy,  which is labelled on the horizontal
      axis with the number given in Table~\ref{galaxy-sample}, we show $\beta$ for 
  $r=1\,$kpc (4 galaxies, red circles), $r=2\,$kpc (15 galaxies, cyan triangles), and
  the total galaxy (33 galaxies, black stars).
  The horizontal lines indicate the median values, red for $r=1\,$kpc,  
    dashed cyan line for $r=2\,$kpc and the dash-dot black line for the total 
  galaxy.}
  \label{distribucion-beta}
\end{figure}

\begin{figure}
  \includegraphics[width=0.45\textwidth]{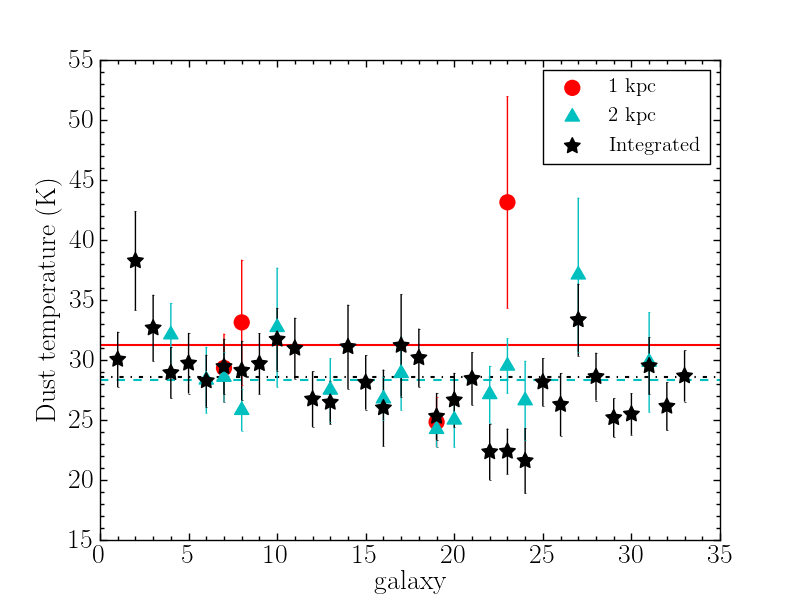}
  \caption{Dust temperature obtained through the grey body fitting
     for each galaxy labelled according to Table~\ref{galaxy-sample}.
      We give the fitted temperature if available for $r=1\,$kpc 
(4 galaxies, red circles), $r=2\,$kpc (15 galaxies, cyan triangles), and
  the total galaxy (33 galaxies, black stars). The horizontal lines 
  indicate the median values for $r=1\,$kpc (solid red line), $r=2\,$kpc (dashed cyan  line)
  and total galaxy (dash-dot black line).}
  \label{distribucion-temperaturas}
\end{figure}

Figure~\ref{exampleSEDs} shows a few examples of the best-fits to
the SEDs for $r=1\,$kpc, $r=2\,$kpc, and the total galaxy. 
In the rest of the galaxies in the sample can be found in the online version.
The values of the best-fit $\beta$ and $T$ parameters are listed in
Table~\ref{tabla-fitting-results}. 
Figure \ref{distribucion-beta} shows 
the $\beta$ distribution for each galaxy for  $r=1\,$kpc, $r=2\,$kpc, and the
total galaxy. In Fig.~\ref{distribucion-temperaturas} we show, for each galaxy, the dust
temperature obtained in the fitting of the different regions (1\,kpc, 2\,kpc,
the total galaxy). 
In Table~\ref{estadistica-ajuste} we provide the statistical information
corresponding to Figs.~\ref{distribucion-beta} and
\ref{distribucion-temperaturas}.
We show in parenthesis the total values for those 14 galaxies in our sample
with SED fits within $r=2\,$kpc so
we can  compare the results for these galaxies in 2 kpc and the total galaxy.

\begin{table*}
 \caption{Results from grey body fits. The full table is
 available in the online version.}
 \label{tabla-fitting-results}

 \begin{tabular}{@{}ccccccccccccccc}
  \hline
  &  \multicolumn{3}{c}{ emission in 1 kpc}&  \multicolumn{3}{c}{ emission in 2 kpc}
  &  \multicolumn{3}{c}{ Integrated emission}  
  & \multicolumn{2}{c}{ Integrated emission with $\beta$=2}& 70 $\mu$m   \\ 
   Galaxy  &  T   &  $\beta$ &   $\chi^2$ &  T   &  $\beta$ 
  &   $\chi^2$ &  T   &  $\beta$ &   $\chi^2$ 
  &    T  &  $\chi^2$ & excess (\%)\\
   
  \hline
  ESO323-G077 &  &  &  &  & &  & 30$\pm$2& 1.7$\pm$0.2& 0.21  & 27.1$\pm$0.6  & 0.90 & 6 \\
  IC5063 &  &  &  &  &  &  & 38$\pm$4& 1.0$\pm$0.2& 2.82   & 25.9$\pm$0.8& 6.99 & 59 \\
  Mrk1066 & &  &  & &  &  & 33$\pm$3& 1.8$\pm$0.2& 0.56   & 30.8$\pm$0.7& 0.56 & 9 \\
  NGC1068 & &  & & 32$\pm$3 & 1.9$\pm$0.2 & 0.39 & 29$\pm$2& 1.9$\pm$0.2& 0.01   & 27.8$\pm$0.6& 0.11 & 1 \\
  NGC1320 & &  &  & &  &  & 30$\pm$3  & 1.5$\pm$0.2& 0.88   & 25.3$\pm$0.6& 1.75 & 22 \\
 
  \hline
  \end{tabular}

\end{table*}

For galaxies with SED fits in the three physical regions, we find that the nuclear
regions with $r=1\,$kpc 
  have the highest temperatures in agreement with the spatially resolved
  maps of the dust temperatures of \cite{SanchezPortal2013} for a few nearby
  Seyfert galaxies. 
  The values obtained for the dust emissivity indices and temperatures for
  the integrated  values are $1.0 <\beta <2.0$ and $21 <T<38\,$K. These  values of 
  temperatures are in agreement with  those obtained by 
  \cite{Dale} ($18 <T<40\,$K) from fits to the $100-500\,\mu$m SEDs of KINGFISH galaxies. 
   Our $\beta$ 
  values are also in agreement with the ones of \cite{Dale} ($1.2 <\beta <1.9$).
  \cite{Perez-garcia} studied 10 Seyfert galaxies observed with {\it ISO} and obtained
  that the  MIR to FIR SEDs can be reproduced with three different components: warm, cold and
  very cold dust. Our temperature range is between the very cold component 
  (T$\sim10-20\,$K) and the cold component (T$\sim40-50\,$K), so it may be due to
  dust heated by stars in the disc (cold component) and by the general interstellar
  radiation field of the galaxy (very cold component).

For  some galaxies (e.g., NGC~3081, NGC~3783,  and NGC~5347)
the fit to the integrated SED does not
  reproduce well the $70\,\mu$m data point. We also tried fits to the integrated SEDs without
  this data point and found that for the majority of galaxies  
the reduced $\chi^2$ values are higher with 70 $\mu$m data than without it. 
This suggests  that in some galaxies this  excess of $70\, \mu$m emission
requires another
component with a higher dust temperature, which could
be associated with dust heated by the AGN.
We will come back to this issue in Section~\ref{Section-criteria}.

\begin{table*}
  \caption{Statistical information for the modified black body fits.}
 \label{estadistica-ajuste}
 \begin{tabular}{lccccc}
  \hline
  Quantity &Region&Number& Mean& $\sigma$ & Median\\
  \hline
  Dust emissivity $\beta$ & 1\,kpc  &  4 &  1.3 &  0.7 &  1.6 \\
  Dust emissivity $\beta$ & 2\,kpc  &  15 &  1.9 &  0.4 &  1.8 \\
  Dust emissivity $\beta$ & integrated & 33 (15)&  1.6 (1.6)&  0.2 (0.2)&  1.6 (1.5)\\
  Reduced $\chi^2$ & 1\,kpc&  4 &  1.47 & 1.34 &  0.77 \\
  Reduced $\chi^2$ & 2\,kpc&  15 & 0.90 & 0.81 &  0.57 \\
  Reduced $\chi^2$ & integrated & 33 (15) &  0.76 (0.55)& 0.66 (0.40)& 0.65 (0.45)\\
  Reduced $\chi^2$, $\beta$=2 & integrated& 33 (15)&  1.37 (1.23)&  1.21 (0.73)& 1.28 (1.28)\\
  Dust temperature $T$(K) &1\,kpc&  4 &   33 &  7 & 31 \\
  Dust temperature $T$(K) & 2\,kpc&  15 &  29 &  39 &  28 \\
  Dust temperature $T$(K) & integrated & 33 (15)&  28 (28)& 3 (3) & 29 (28)\\
  Dust temperature $T$(K), $\beta$=2 & integrated & 33 (15)&  25 (24)& 3 (3) & 26 (25)\\
  \hline
 \end{tabular}

 Notes.--- We show in parenthesis the total values for the  15 galaxies that also have SED fits within
 $r=2\,$kpc.
\end{table*}

\begin{figure*}
  \begin{minipage}{180mm}
    \begin{center}

      \includegraphics[width=43mm]
      {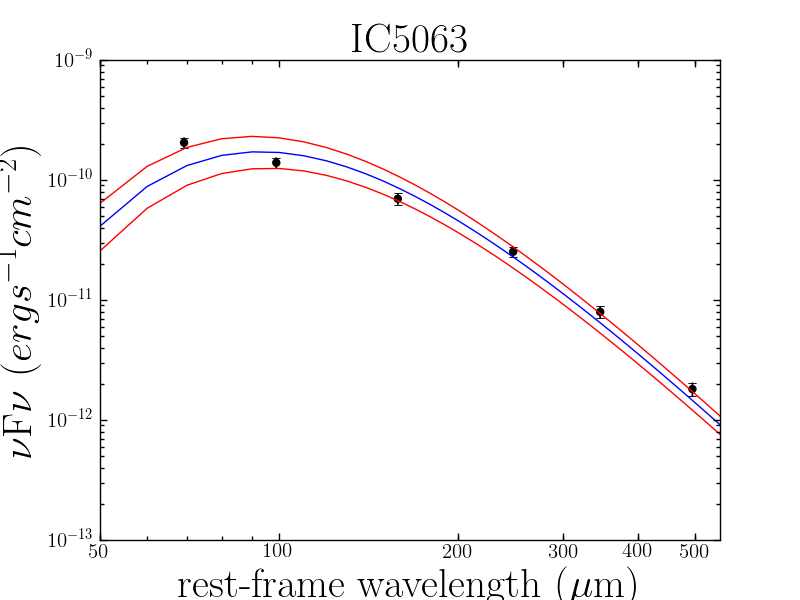}{\vspace{0cm}}
      \includegraphics[width=43mm]
      {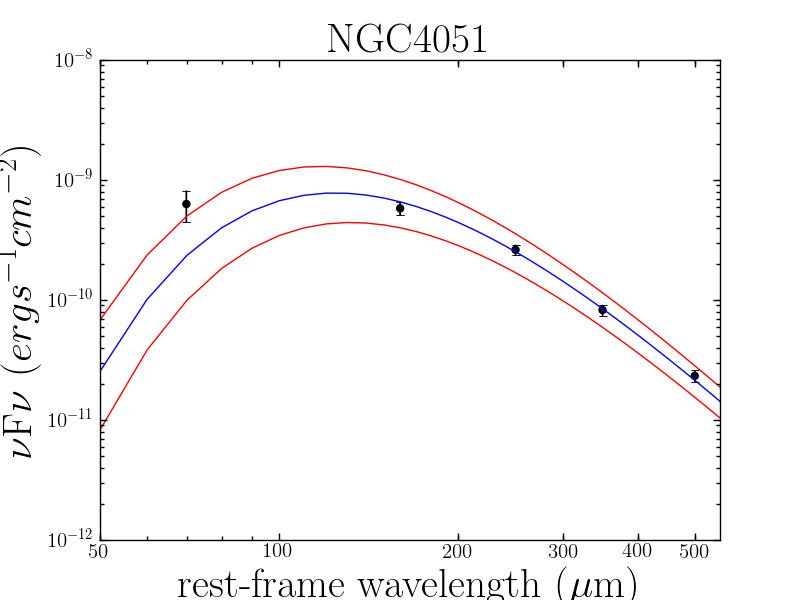}{\vspace{0cm}}
      \includegraphics[width=43mm]
      {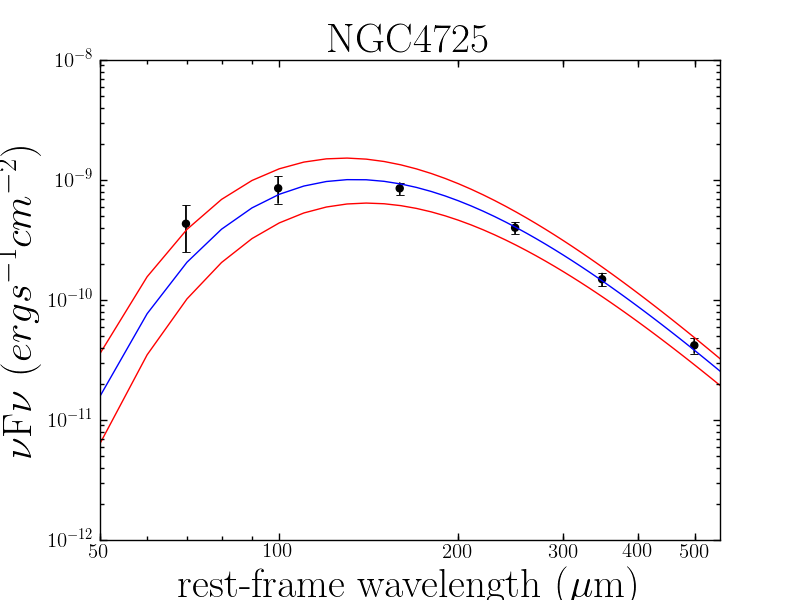}{\vspace{0cm}}
      \includegraphics[width=43mm]
      {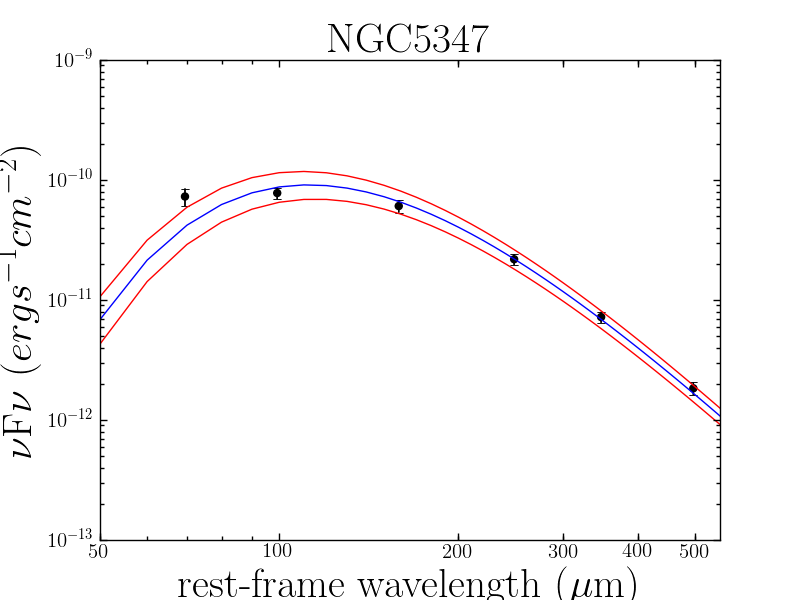}{\vspace{0cm}}

      \caption{Fits to the integrated SEDs with $\beta$=2 for those
        RSA Seyfert galaxies with a clear excess over the fit
        at $70\,\mu$m. The blue line indicates 
        the best fit, whereas the red lines delineate the best fit $\pm 1\sigma$ uncertainty.
      The fits for the rest of the RSA Seyferts in our sample are in the online version.}
      \label{figuras-exceso-SEDS}

    \end{center}
 
  \end{minipage}

\end{figure*}

To  quantify the 70 $\mu$m excess in the integrated SED from the single
temperature fits, we performed  new 
grey body fits imposing $\beta$=2, which is the typical value for star forming galaxies 
\citep{Li-Draine}. The statistical information about the fitted values of the dust temperature for $\beta=2$ are listed in
Table~\ref{estadistica-ajuste}.
The range of dust temperatures obtained with the fit imposing $\beta$=2
 is $18-31\,$K, which are the normal temperatures for star forming galaxies.
As expected, since $\beta$ and $T$ are anticorrelated \citep[see e.g.][and also
  Figs.~\ref{distribucion-beta} and \ref{distribucion-temperaturas}]{Galametz2012},
for a given galaxy the fits with fixed dust emissivity $\beta=2$ produce
lower dust temperatures.

In Fig.~\ref{figuras-exceso-SEDS} we show 4 RSA Seyfert galaxies with integrated SED
fits with fixed dust emissivity $\beta=2$ showing  the highest excesses at $70\,\mu$m
over the single temperature  fits, which puts them
  clearly above the $1\sigma$ uncertainty of their fits. The data are shown in black, and
  the best fit with $\beta$=2 is shown 
in blue. The red lines indicate the best fit $\pm 1 \sigma$ of the free 
parameters.
The results for these fits for the rest of the sample are in the online version.
 
\begin{figure}
  \includegraphics[width=0.45\textwidth]{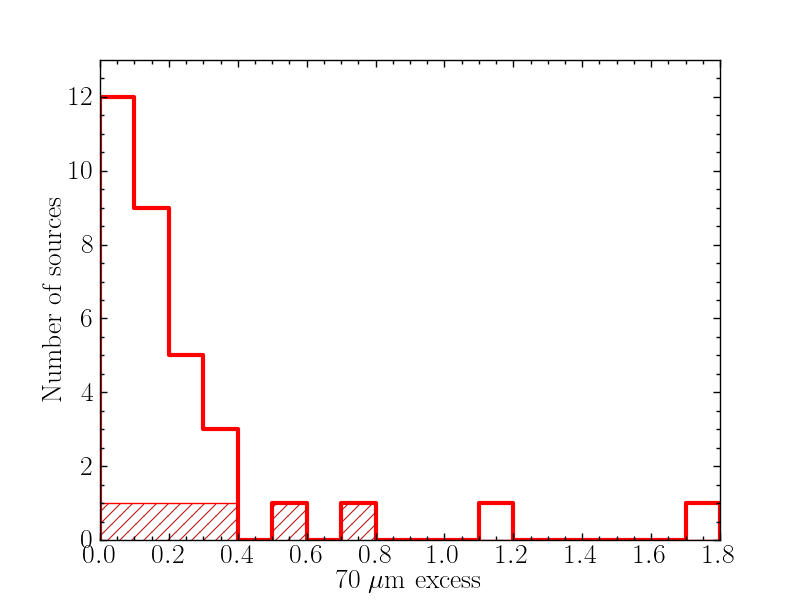}
  \caption{Distribution of the 70 $\mu$m integrated flux excesses calculated as 
    $\frac{f_\nu({\rm obs})-f_\nu({\rm model})}{f_\nu({\rm model})}$ (see
    Section~\ref{sec:grey body-fits}).  The hatched histogram
  shows the final sample of Seyferts with 
   significant nuclear $70\,\mu$m AGN emission
  (see Section \ref{sec:final-candidates}).
}
  \label{distribucion-exceso70micras}
\end{figure}

 We quantify the
  excess over the fit at $70\,\mu$m as,
  $\frac{f_\nu({\rm obs})-f_\nu({\rm model})}{f_\nu({\rm model})}$, where $f_\nu({\rm model})$
is the grey body fitted with $\beta=2$ and free dust temperature and $f_\nu({\rm obs})$
is the observed integrated value at $70\,\mu$m. 
In  Table~\ref{tabla-fitting-results} we list in
the last column the 70 $\mu$m excess. 
The average  and median $70\,\mu$m excess of the  sample are  26 per cent and 16 per cent, respectively,
although some galaxies have  high excesses, such as NGC~4051 (170 per cent excess) and 
NGC~4725 (114 per cent excess). In Fig.~\ref{distribucion-exceso70micras} we show 
the 70 $\mu$m excess distribution. We note that all the sources have
a positive excess. These $70\,\mu$m excesses might be due to a hotter dust component
not necessarily related to
dust heated by an AGN
\citep[][]{Galametz2012,Almudena}. We will discuss this further in
Section~\ref{Section-criteria}.

\subsection{Star Formation Rates}
\label{subsection-SFR}

In this subsection we calculate the nuclear and extranuclear 
SFR using the $70\,\mu$m luminosity  and  compare them with the SFR obtained
  with other indicators. The goal is to identify galaxies where there is 
  excess emission at $70\,\mu$m due to the AGN. We use the 
recipe from \citet{Li2013}, which assumes a \citet{Kroupa} IMF:

\begin{eqnarray}
SFR (70~\mu m) (M_{\sun}~yr^{-1}) = C_{70~\mu{\rm m},{\rm region}} \times 10^{-43} \nonumber \\
\times L(70~\mu m)(erg~s^{-1})
\end{eqnarray}

The calibration coefficient $C_{70~\mu{\rm m},{\rm region}}$ is different
depending on the region of the galaxy.
To calculate the SFR we used the  coefficients from
\citet{Calzetti} and \citet{Li2010}, which are $C_{70~\mu {\rm m,galaxy}}= 0.58$ and 
$C_{70~\mu {\rm m,0.5-2~kpc}}= 0.94$, respectively. The \citet{Calzetti} coefficient was derived from the integrated
emission of galaxies and includes contributions
from the diffuse component at $70\,\mu$m whereas the \citet{Li2010} measurements
are local (star forming regions on scales of
 $0.5-2\,$kpc) and  minimize the contribution of any diffuse component.

\begin{figure}
  \includegraphics[width=0.5\textwidth]{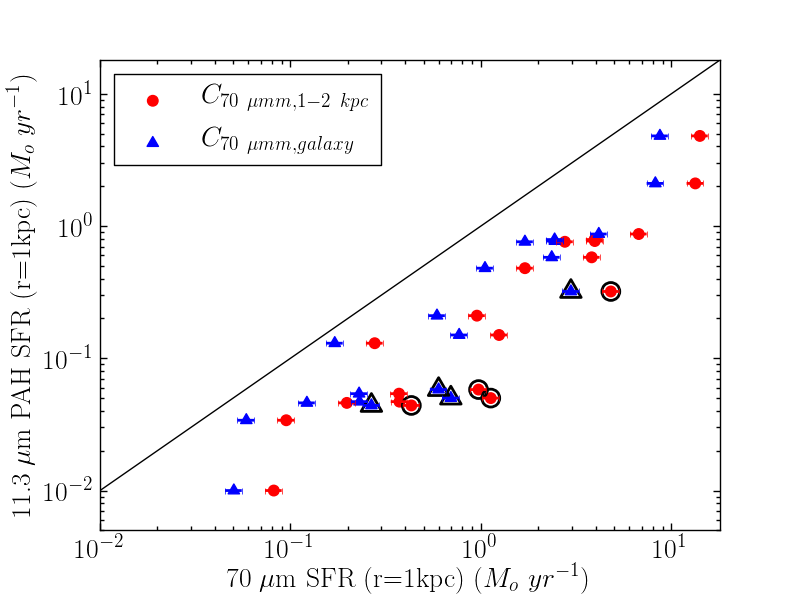}
  \caption{Nuclear SFR ($r=1\,$kpc) from the $70\,\mu$m luminosity versus
  the 11.3 $\micron$ PAH SFR ($r=1\,$kpc)
    from  DSR2012. Red circles are the $70\,\mu$m-based SFR 
    calculated with  $C_{70\mu {\rm m, 0.5-2kpc}}= 0.94$ (regions 
        $0.5-2$\,kpc in size),
      whereas the blue triangles  are SFR obtained with $C_{70\mu {\rm m,galaxy}}=  0.58$ (entire
  galaxy). The black line indicates
  the  1:1 relation. The $70\,\mu$m nuclear SFR uncertainties are
  derived by propagation of errors in the  
  SFR formula, using the flux errors given in Tables~\ref{tabla-fotometria-1kpc}, \ref{tabla-fotometria-2kpc}
  and \ref{tabla-fotometria-total}.
  The black symbols show the final selected Seyferts with  
   significant nuclear $70\,\mu$m  AGN emission
    (see Section~\ref{sec:final-candidates}). We note that there are two symbols
    for each galaxy, one for each coefficient.}
  \label{SFR-1kpc}
\end{figure}

\begin{figure}
   \includegraphics[width=0.5\textwidth]{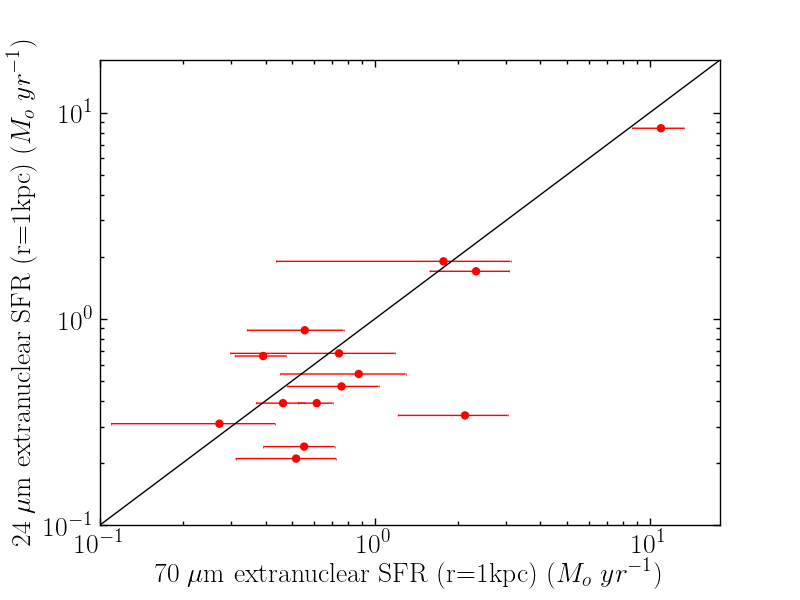}
    \caption{Extranuclear SFR  ($r>1\,$kpc) estimated with the $70\,\mu$m-based SFR  versus extranuclear
       ($r>1\,$kpc) 24 $\micron$ SFR  from  DSR2012.
      The black line indicates the 1:1 relation. The 70 $\mu$m extranuclear SFR uncertainties are 
      calculated by propagation of errors in the 
      SFR formula, using the flux errors given in Tables~\ref{tabla-fotometria-1kpc}, \ref{tabla-fotometria-2kpc}
  and \ref{tabla-fotometria-total}.} 
  \label{SFR-extranuclear}
\end{figure}

Figures~\ref{SFR-1kpc} and \ref{SFR-extranuclear} compare the values for the 
SFR obtained by  DSR2012
with our $70\,\mu$m-based SFRs for  the nuclear  ($r=1\,$kpc) and
extranuclear  ($r>1\,$kpc) regions, respectively. We have 20 galaxies in common with  DSR2012
with nuclear SFR values and 14 galaxies  in common with the  DSR2012
extranuclear SFR values. They obtained their
 nuclear SFR values from the luminosity of the $11.3\,\mu$m  PAH feature and the extranuclear 
 SFR from the luminosity of the  extended 
$24\,\mu$m continuum emission after subtracting the contribution from the 
nuclear source. 
For the nuclear SFRs our values derived with the $C_{70~\mu{\rm m},{\rm galaxy}}$ coefficient for the
whole galaxy are 
on average four times higher
than those of  DSR2012,  and even a factor of  7 if we used
  the coefficient for $0.5-2\,$kpc regions. These differences are larger than the
quoted scatter of 0.14dex of the \cite{Li2010} calibration with some galaxies showing large 
discrepancies. For instance, NGC~3783 has a nuclear $70\,\mu$m based SFR 13.6 times higher
than the nuclear $11.3\,\mu$m PAH based SFR. For the extranuclear SFRs,
the majority of the galaxies in our sample have  higher 
SFRs using the 70 $\mu$m emission than the  DSR2012 value, although
the galaxies appear to be closer to the 1:1 line.

The discrepancies in SFRs found for the RSA Seyferts are similar to those
for the {\it Swift}/BAT AGN of \cite{Mushotzky}. They calculated the total SFR 
using the \citet{Calzetti} $70\,\mu$m calibration,  that is for the entire galaxy,
and compared them with the SFR obtained from the 
 $11.3\,\mu$m PAH feature using the calibration from
 DSR2012. They obtained a discrepancy of  a factor of 3 between 
the two SFR. \citet{Petric} also found that the integrated galaxy SFRs of a sample of
PG quasars obtained from the $11.3\,\mu$m 
PAH feature were on average 3 times lower than the SFR estimated from the   $40-500\,\mu$m
emission.

 All the differences in SFR described above could be explained if 
the carriers of the $11.3\,\mu$m PAH feature were being destroyed by the 
AGN radiation field, if there were aperture
correction issues and/or if there were systematics in the calibrations.
  DSR2012 argued that SFR estimates using the $11.3\,\mu$m PAH feature appear to be
  robust to the effects of AGN and shock processing based on the good correlation with the
  [Ne\,{\sc ii}]$12.8\,\mu$m line on kpc scales. Additionally \cite{Esquej2014} showed that even
  on smaller physical nuclear scales of nearby Seyferts there is no strong
  evidence for destruction of the $11.3\,\mu$m PAH carriers.
The SFR differences do not seem to be due 
to the aperture corrections applied because all three works (DSR2012,
\citealt{Mushotzky} and \citealt{Petric}) also applied aperture corrections to their fluxes.

\section{Identifying galaxies with significant
$70\,\mu$m emission due to AGN heated dust }
\label{Section-criteria}

As the goal of this work is to select galaxies whose $70\,\mu$m 
emission is mostly due to dust heated by the AGN, in this section 
we put forward four different criteria to identify this type of galaxies. 
  We also compare
them with other results in the literature and 
  propose candidate RSA galaxies with 
  significant nuclear $70\,\mu$m AGN emission.

 \subsection{Elevated  $f_\nu(70\,\mu$m)/$f_\nu(160\,\mu$m) flux ratios}\label{subsection-flux-ratios}
 
 If the $f_\nu(70\,\mu$m)/$f_\nu(160\,\mu$m) flux ratios are higher than the typical
 values for star forming galaxies, this  might indicate 
 that part of the nuclear 70 $\mu$m emission is due to the 
 dust heated by the AGN instead of star formation. To select
 the galaxies with an elevated  $f_\nu(70\,\mu$m)/$f_\nu(160\,\mu$m) flux ratio
 we choose all the 
 galaxies with a value higher than the median plus  $1.4826 \times {\rm M.A.D}$, where the M.A.D 
 (absolute median deviation) is
 calculated as the median of the absolute deviations from the 
 data median, ${\rm M.A.D.} ={\rm median}(|x_i - {\rm median_{data}}|)$. We do this for 
 the $r=1\,$kpc, $r=2\,$kpc, and integrated flux ratios using the
   statistics in Table~\ref{estadistica-colores}, each one with its
   own value.

   We select  9 galaxies with this criterion (see Table~\ref{tabla-criterios}).
   We note that the 2 galaxies with high $f_\nu(70\,\mu$m)/$f_\nu(160\,\mu$m) 
   flux ratios within $r=1\,$kpc also show it within $r= 2\,$kpc. All the galaxies 
   with a high flux ratio within $r=2\,$kpc have high flux ratio at $r=1\,$kpc or have no
   measurements within 1 kpc. The same happens with the 
   total flux ratio, galaxies with high total flux ratio have also
   high 2 kpc flux ratio or have no measurement at 2 kpc. However, not all
   galaxies with a high nuclear
   flux ratio have also high total flux ratio.  Only NGC~5506 is
   selected from the flux ratios within
   $r=2\,$kpc and with
   the total galaxy flux. The rest have extended diffuse emission at 160 $\mu$m and therefore
   their measured integrated flux is higher at 160 $\mu$m than at 70 $\mu$m (see
   Fig.~\ref{ExampleHerschelimages} and Table~\ref{tabla-fotometria-total}).

\begin{figure*}
  \begin{minipage}{180mm}
    \begin{center}
      
      \includegraphics[width=0.48\textwidth]{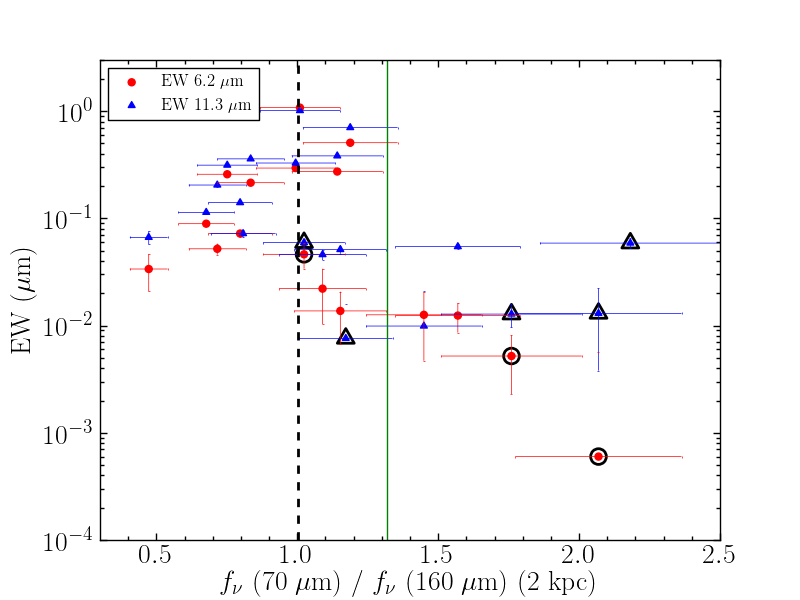}{\vspace{0cm}}
      \includegraphics[width=0.48\textwidth]{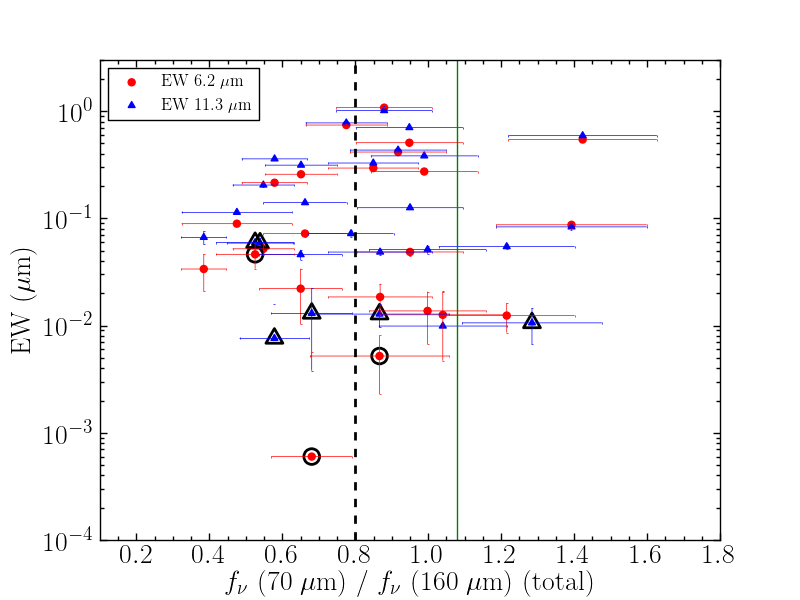}{\vspace{0cm}}
  
      \caption{Relation between the $f_\nu(70\,\mu$m)/$f_\nu(160\,\mu$m) flux ratio and the EW 
        of the $6.2\,\mu$m (red circles) and $11.3\,\mu$m (blue triangles)  PAH features as measured
        from {\it Spitzer}/IRS spectra.
        The left panel shows the relation
  for $r=2\,$kpc and the right panel for the total galaxy. The dashed black line
  indicates the median of each sample and the green line the  median plus  1.4826 $\times$ M.A.D.  The black symbols show the final selected Seyferts
  with   significant nuclear  $70\,\mu$m AGN emission   (see Section~\ref{sec:final-candidates}). }

  \label{colores-vs-EW}

  \end{center}
  \end{minipage}
\end{figure*}

 Figure~\ref{colores-vs-EW} shows the relation between the 
 $f_\nu(70\,\mu$m)/$f_\nu(160\,\mu$m) flux ratio and the EW of the 6.2 and $11.3\,\mu$m
 PAH features measured from {\it Spitzer}/IRS spectra. The emission from these features is
 an indicator of the presence of on-going/recent star formation
 activity \citep{Peeters2004} but  can be diluted by the AGN continuum,
 resulting in lower EW  \citep{Spoon2007,Diaz-Santos2010,HernanCaballero2011,Esquej2014}. 
 Note that not all the galaxies in our sample have a value of 
 the EW (see Table \ref{galaxy-sample}). There is one galaxy that satisfies the criterion
 for $r=2\,$kpc without EW data.

 As expected, those galaxies with a high value of the $f_\nu(70\,\mu$m)/$f_\nu(160\,\mu$m) 
 have a small value of the EW of the  PAH features, indicating that it is a good criterion to
 select galaxies with the dust heated by the AGN. 
  We note that the size of the IRS SL slit is similar to the FWHM at 70 $\micron$.
 The only discrepant galaxy  is
 the Seyfert 2 galaxy
 Mrk~1066, which has a high total $f_\nu(70\,\mu$m)/$f_\nu(160\,\mu$m) flux ratio,
   but also a high value of the EW of the PAHs. \cite{AAH2014} and \cite{RamosAlmeida2014} have
   shown, based on high angular resolution ($0.3$arcsec) MIR spectroscopy, that this
   galaxy has a strong nuclear
   starburst including  the central $\sim 60\,$pc region. This would explain
   the elevated FIR colours of
   this galaxy.  However, as can be seen from Fig.~\ref{colores-vs-EW},  not all galaxies
     with low  PAH EWs
     features satisfy this criterion. Since the PAH features probe mostly the
     emission from O and B stars \citep{Peeters2004}, it is possible that some galaxies
     in our sample have most of
     their FIR emission
      mainly due to heating from older stars \citep[see e.g.][]{Li2010}
     explaining why they do not have bright PAH
     emission but still have  normal FIR colours. Alternatively for 
     these galaxies the torus emission  might dominate in the MIR but it decays rapidly in the
     FIR (see \citealt{Mullaney2011}). 
 
 \subsection{Dust temperature gradient higher than typical star forming galaxies}\label{subsection-T-gradient}
 
 If the nuclear dust temperature is higher than those typical of
 star forming galaxies, this may indicate that the nuclear dust is 
 not only heated by star forming but also by the AGN.  However, because
   star-forming galaxies show a range of dust temperatures \citep[][]{Dale}, for a given
   galaxy we compare the nuclear ($r=1\,$kpc and $r=2\,$kpc) temperatures to
   the dust temperature fitted for entire galaxy.
   
   To select
 the galaxies with an elevated  dust temperature gradient we use as criterion the ratio
 between the nuclear temperature and the total temperature for each galaxy.
 We select all the 
 galaxies with a value higher than the median plus 1.4826 $\times$ M.A.D  
 for our sample of galaxies.
 For this criterion we can use  15 galaxies, which are the galaxies with
 at least four data points to fit the $r=2\,$kpc SEDs. These  15 galaxies have
 similar mean integrated dust temperatures to those of the other RSA Seyfers
 (see Table~\ref{estadistica-ajuste}).
 With this criterion we select 3 galaxies with higher dust temperature gradient than the 
 typical in our sample.  These are NGC~4579, NGC~4594, and NGC~4725.

\begin{table*}
 \caption{  Summary of criteria to select galaxies with a significant AGN contribution at $70\,\mu$m.
 }
 
 \begin{tabular}{@{}lccccccccccc}
  \hline
  Galaxy & \multicolumn{3}{c}{\#1} & \multicolumn{2}{c}{\#2} & \#3 & \#4\\ 
  & \multicolumn{3}{c}{$f_{\nu}(70\,\mu{\rm m})/f_{\nu}(160\,\mu{\rm m})$}
  & \multicolumn{2}{c}{$T_{\rm region}/T_{\rm Integrated}$}
  & 70 $\mu$m  & SFR$_{70\mu{\rm m}}$/SFR$_{\rm D-S}$ &  criteria \\
  & $r=1\,$kpc & $r=2\,$kpc & Integrated & $r=1\,$kpc & $r=2\,$kpc & excess & $r=1\,$kpc \\ 
  \hline 
 ESO~323-G077 & -  & - &x& - & -  & x& - & 0/2\\
  {\bf IC~5063} & - & - &\checkmark& - & - &  \checkmark & - & 2/2\\
 {\bf Mrk~1066} & - & - &\checkmark& - & - & x & - & 1/2\\
  NGC~1068 & \checkmark & \checkmark &x& - & x  & x & - & 1/3\\
  NGC~1320 & - & x &x& - & -  & x &- & 0/2\\
  NGC~1365 & x & x&x & - & x  & x & x & 0/4\\
  NGC~1386 & x & x &x& x & x  & x & x & 0/4\\
  NGC~1808 & x & x &x& x & x  & x & - & 0/3\\
  NGC~2110 & - & x &x& - & -   & x & - & 0/2\\
  NGC~2273 & - & x &x& - & x   & x & x & 0/4\\
  NGC~2992 & - & x &x& - & -   & x & x & 0/3\\
  NGC~3081 & - & x &x& - & -   & \checkmark &x  & 1/3\\
  NGC~3227 & x & x &x& - & x  & x & x & 0/4\\
  NGC~3281 & - & \checkmark& x & - & -   & x & x & 1/3\\
  {\bf NGC~3783 }& - & \checkmark &x& - & -   & x &\checkmark  & 2/3\\
  NGC~4051 & x & x &x& - & x  & \checkmark &x  &  1/4\\
  {\bf NGC~4151} & \checkmark & \checkmark&x & - & x  & x & \checkmark & 2/4\\
  {\bf NGC~4253} & - & -&\checkmark & - & - & x & - & 1/2\\
  NGC~4258 & x & x &x& x & x &  x &x & 0/4\\
  NGC~4388 & - & x&x & - & x &  x & x & 0/4\\
  NGC~4507 & - & - &x& - & -  & x & - &  0/2\\
  {\bf NGC~4579} & x & x &x& - & \checkmark   & \checkmark &x  & 2/4\\
  NGC~4594 & x & x &x& \checkmark & \checkmark  & x & x & 1/4\\
 {\bf NGC~4725} & - & x &x& - & \checkmark   & \checkmark & x  & 2/4\\
  NGC~5135 & - & - &x& - & -  & x & - &  0/2\\
  {\bf NGC~5347} & - & x &x& - & -   & \checkmark &  - & 1/2\\
  NGC~5506 & - & \checkmark & \checkmark& - & x   & x & x  & 1/4\\
  NGC~7130 & - & - &x& - & -  & x & - &  0/2\\
  NGC~7172 & - & x &x& - & -  & x & x &  0/3\\
  {\bf NGC~7213} & - & \checkmark &x& - & -  & x & \checkmark & 2/3\\
  NGC~7465 & - & x &x& - & x  & x & - & 0/3\\
  {\bf NGC~7479} & - & x &x& - & -  & \checkmark & \checkmark & 2/3\\
  NGC~7582 & x & x &x& - & -  & x & x & 0/3\\

  \hline
  
 \end{tabular}
 
 Notes.--- In bold are marked galaxies satisfying at least half of the conditions.
 
 \label{tabla-criterios}
\end{table*}

 \subsection{Excess 70 $\mu$m emission with respect to the fit of the FIR 
  SEDs with a grey body}\label{subsection-excesos}
  
  The $70\,\mu$m excess with respect to the fit of the FIR 
  SEDs with a grey body with $\beta$=2  could in principle indicate that this emission is 
  not only due to  star formation but that there is some contribution from the AGN.  We select 
  those galaxies whose excess at $70\,\mu$m is 
  higher than the median plus  1.4826 $\times$ M.A.D
   for our sample of galaxies (28 per cent).
  
\begin{figure}
  \includegraphics[width=0.5\textwidth]{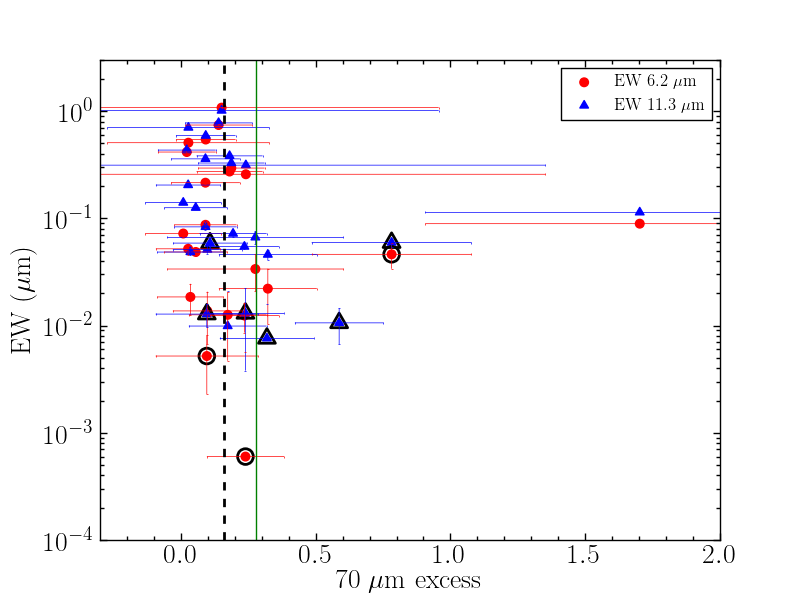}
  \caption{Relation between the 70 $\mu$m excess and the EW 
    of the $6.2\,\mu$m (red circles) and $11.3\,\mu$m (blue triangles) PAH features as measured
        from {\it Spitzer}/IRS spectra.
     The dashed black line
  indicates the median of each sample and the green line the  median plus  1.4826 $\times$ M.A.D.
    The black symbols show the final selected Seyferts with  
    significant nuclear $70\,\mu$m AGN emission
  (see Section~\ref{sec:final-candidates}).}
  \label{exceso-vs-EW}
\end{figure}

  Figure~\ref{exceso-vs-EW} plots the $70\,\mu$m excess against the EW of the 6.2 and $11.3\,\mu$m  PAH
  features as measured from the {\it Spitzer}/IRS spectra.
  Since those galaxies with a considerable excess also  have small values of EW
  for both features, it is likely that this criterion selects galaxies with a
  contribution to the $70\,\mu$m from dust heated by the AGN. Again, as found
    for the FIR colours, not all galaxies with low EW of the PAH features have a $70\,\mu$m
    excess.
  
  \subsection{  Excess of nuclear SFR obtained from 
$70\,\mu$m over SFR from mid-infrared indicators}\label{subsection-excesos-SFR}
    
  As showed in Section~\ref{subsection-SFR}, all the nuclear $70\,\mu$m-based SFR are
  systematically higher than those obtained by 
   DSR2012 using the $11.3\,\mu$m PAH feature luminosity.
  We explained this as due to systematics in the calibrations. For this reason,
  we select those galaxies with the most discrepant
  values of SFR,  i.e. all the 
 galaxies with a value of $SFR_{1kpc}/SFR_{D-S}$ higher than the median plus
   1.4826 $\times$ M.A.D.. This could indicate that part of the nuclear 70 $\mu$m emission
 of these galaxies is due to the dust heated by the AGN. For this criterion
 we can use 20 galaxies  that are the 
 ones in common with  DSR2012 and with measurements at 70 $\mu$m
 for $r=1\,$kpc.  We select  4 galaxies with $70\,\mu$m-based nuclear
 SFR higher than expected taking into account the 
 systematics in the calibrations,  namely, NGC~3783, NGC~4151, NGC~7213, and NGC~7479. These four
 galaxies also satisfy some of the other criteria above.
  
\subsection{Comparison with other works}

The results about the nuclear 70 $\micron$ emission
obtained with the above criteria are in good agreement with the detailed
{\it Herschel} imaging studies 
of the infrared emission of three Seyfert galaxies, NGC~1365, NGC~2992, and NGC~3081 mentioned in
Section~\ref{Section-introduction}.
\cite{Almudena} and \cite{Ismael} found that the AGN emission does not dominate
the $70\,\mu$m emission of  NGC~1365 and NGC~2992, respectively. These two 
galaxies do not satisfy any of our selection criteria. On the other hand,
\cite{Cristina} assumed that  the nuclear 70$\mu$m flux of NGC~3081
is due to dust heated in the torus surrounding the AGN. They modelled the unresolved infrared
emission of this galaxy
with the \cite{Nenkova2008} clumpy torus models and were able
to reproduce the AGN bolometric luminosity. This galaxy satisfies the 70 $\mu$m excess
emission with respect to the fit of the FIR SEDs with a grey body with $\beta=2$ criterion.
The reason why this galaxy does not meet the other criteria may be due to the 
nuclear ring of 2\,kpc in diameter. Because of its distance we were not able to
derive nuclear $r=1\,$kpc dust temperature and $f_\nu(70~\mu m)/f_\nu(160~\mu m)$ flux ratio.
Therefore we were restricted to measurements within $r=\,2$ kpc and
the total galaxy.

 We have 4 galaxies in common  with the work of 
\cite{Mullaney2011}, namely,
NGC~2110,  NGC~4507, NGC~5506, and IC~5063. Among these they found that 
the only galaxy dominated by the AGN emission at $60\,\mu$m is IC~5063 in good agreement
with our results.

\subsection{Inspection of the candidates to significant nuclear 
$70\,\mu$m AGN emission}\label{sec:final-candidates}

Table~\ref{tabla-criterios} summarizes  the selection criteria fulfilled by each galaxy.
 In the last column we indicate
the number of criteria satisfied by each galaxy. 16 galaxies 
(48 per cent of the sample) satisfy at least one of these  requirements, while  10 of them fit at least
half of the  criteria. 
We found no differences between Seyfert 1 and 2 in terms of the selection  requirements (p=0.30 with the Fisher's test). Of the 
 16 galaxies that satisfy at least one criterion,  9 are Seyfert 1 and 7 Seyfert 2.
We also found no differences between the galaxies that satisfy at least one criterion
and the galaxies that donot satisfy any  of them in terms of AGN luminosity or 70 $\mu$m 
luminosity.

 In what follows we take a conservative approach by only considering the
   10 RSA Seyferts in our sample that satisfy
  half or more of the  criteria defined in the previous sections to select
  candidates with a significant AGN contribution to the nuclear $70\,\mu$m
  emission.

  We start by discarding two galaxies, NGC~4253 and Mrk~1066, as both
  show  $11.3\,\mu$m PAH emission in the inner 0.5\,arcseconds, equivalent to physical
  sizes of 120 and 145\,pc, respectively \citep{AAH2014, AAH2015} as well as high SFR within $r=1\,$kpc scales.
   This probably suggests that the elevated
  $f_\nu(70\,\mu$m)/$f_\nu(160\,\mu$m) flux ratios are due to strong star formation rather than
  AGN dominated fluxes at $70\,\mu$m \citep[see][for a detailed discussion of Mrk~1066]{RamosAlmeida2014}.

  We next discuss NGC~4579 and NGC~4725, which are among the least luminous AGN in our sample of
  RSA Seyferts. Both are close to
  the low-luminosity AGN (LLAGN) limit of $10^{42}\,{\rm erg\,s}^{-1}$ below which the dusty torus
  is predicted to disappear
  \citep{Elitzur2006}. As can be seen from Table~\ref{tabla-fitting-results}, both galaxies
  show a
  significant excess at $70\,\mu$m with respect to the $\beta=2$ grey body fit to the integrated SED.
  However, if we assumed that the excess is entirely due to dust heated by the AGN, then the AGN flux at
  $70\,\mu$m would be similar to the measured nuclear $r=1\,$kpc flux for NGC~4579 which is
  resolved (FWHM=600\,pc) at this wavelength. For NGC~4725 the predicted AGN flux would be more than   twenty
  times brighter than that arising from the nuclear (resolved) region with a 1.4\,kpc size (FWHM). We
  therefore conclude that the $70\,\mu$m nuclear emission of NGC~4579 and NGC~4725 at the {\it Herschel}
  resolution is not dominated by
  dust heated from the AGN.

  We are left with  six (18 per cent of the sample) {\it bona fide} candidates,
  namely,  IC~5063, NGC~3783,  NGC~4151, NGC~5347, NGC~7213, and NGC~7479.

  \subsection{MIR and FIR emission of the candidates to significant nuclear $70\,\mu$m AGN emission}\label{sec:six-candidates}
  
  None of the six candidates shows
  $11.3\,\mu$m PAH emission in high angular resolution (0.5-0.7\,arcsec
  scales) MIR spectroscopy
  \citep{AAH2011,GonzalezMartin2013,Esquej2014,AAH2015} and all of them show low values of the nuclear
  SFRs (see Table~\ref{galaxy-sample}).  \cite{Antonio2015} performed a spectral 
  decomposition of 118 {\it Spitzer} IRS spectra of local AGN. The 6 candidates have a high
  AGN contribution (within the IRS slit) at $\lambda<15\,\mu$m and an AGN 
  12 $\micron$ luminosity in agreement with the estimates from nuclear high angular
  resolution spectra {\citep{AAH2011,GonzalezMartin2013}.
  We show the selected galaxies as the hatched histogram in Fig.~\ref{distribucion-exceso70micras} 
  and with black symbols in Figs.~\ref{SFR-1kpc}, \ref{colores-vs-EW}, and \ref{exceso-vs-EW}.
  
   We used the different criteria to estimate the range of the AGN flux at 70 $\micron$.
  For each galaxy we only used the criteria satisfied 
   (see Table \ref{tabla-criterios}). For the galaxies
  that have a 70 $\micron$ excess with respect to the fit of the FIR SEDs with a $\beta=2$ grey body
  (see Section \ref{subsection-excesos}), 
  we can estimate directly the AGN flux. For the galaxies that
  satisfied the elevated  $f_\nu(70\,\mu$m)/$f_\nu(160\,\mu$m) flux ratios (see Section \ref{subsection-flux-ratios}),
  or the excess of nuclear SFR obtained from 
$70\,\mu$m over SFR from mid-infrared indicators (see Section \ref{subsection-excesos-SFR}) we 
make the observed values compatible with the typical values of the sample, and the excess is
assumed to be the AGN emission. For IC~5063 we obtain an AGN flux
at 70 $\micron$ of 1.8 Jy
(40 per cent of the nuclear r = 2 kpc flux) using the two criteria.
For NGC~5347 we can only calculate the contribution with the excess
with respect to the $\beta=2$ grey body fit. 
The AGN contribution is approximately 0.7 Jy (61 per cent of the nuclear r = 1 kpc flux).
For the rest of the galaxies we obtain different fluxes depending on the criterion used to estimate
the 70 $\micron$ AGN flux. The ranges are $1.0-1.3$ Jy for NGC~3783 
($56-73$ per cent of the nuclear r = 1 kpc flux);
$2.4-3.0$ Jy for NGC~4151 ($49-62$ per cent of the nuclear r = 1 kpc flux); 
$0.5-1.1$ Jy for NGC~7213 ($35-76$ per cent of the nuclear r = 1 kpc flux); and
$4.1-5.4$ Jy for NGC~7479 ($43-57$ per cent of the nuclear r = 1 kpc flux).
 We finally note that the estimated AGN $70\,\mu$m fluxes for IC~5063 and NGC~4151 are
in good agreement with the predicted torus FIR emission 
from the extrapolation of the fits to the unresolved
near-infrared and MIR emission \citep{AAH2011,Ichikawa2015} using clumpy torus models.

  Figure \ref{espectro-candidatos} shows the {\it Spitzer}/IRS SL+LL
  spectra for these six galaxies,
  normalized at 30 $\mu$m.
  We also plot the estimated AGN $70\,\mu$m flux ranges
  and the average SEDs of the low luminosity (log($\frac{L_{2-10 keV}}{{\rm erg\,s}^{-1}}$)~<~42.9)
  and high luminosity (log($\frac{L_{2-10 keV}}{{\rm erg\,s}^{-1}}$)~>~42.9)
  AGN of \cite{Mullaney2011}, all of them normalized at 30 $\mu$m.
  All our candidates have MIR and 70 $\micron$ AGN emission entirely
  consistent with the empirically determined low and high 
  luminosity AGN templates of \cite{Mullaney2011}, except around the 
  9.7 $\micron$ silicate feature range for the two most extreme features (NGC~7213 and NGC~7479).

  \begin{figure}
  \includegraphics[width=0.5\textwidth]{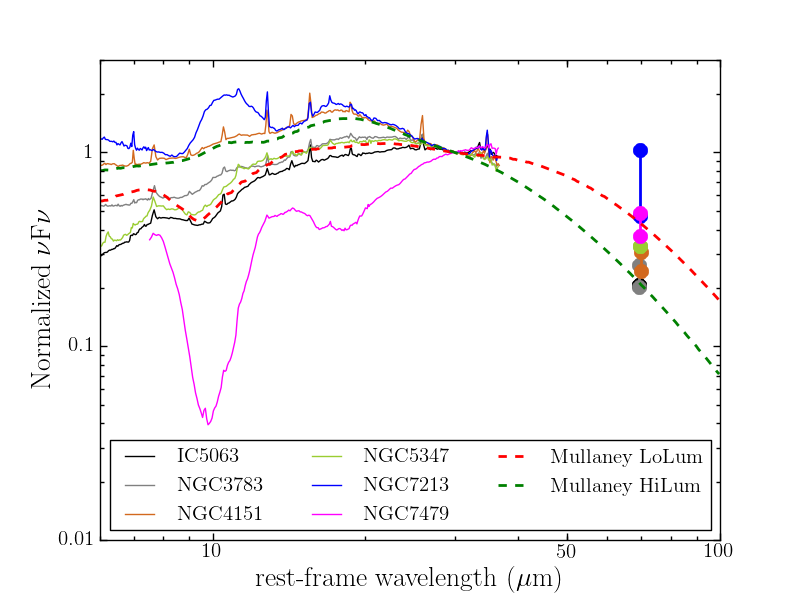}
  \caption{ {\it Spitzer}/IRS SL+LL spectra normalized at 30 $\mu$m of the six candidates to  significant nuclear $70\,\mu$m AGN 
  emission. The  estimated AGN $70\,\mu$m flux ranges (see Section \ref{sec:final-candidates}) 
  are shown with filled circles, normalized to the IRS flux at 30 $\mu$m for each galaxy. The 
  red and green dashed lines indicate, respectively, the average SED of the 
  empirically determined low luminosity  (log($\frac{L_{2-10 keV}}{{\rm erg\,s}^{-1}}$)~<~42.9)
  and high luminosity  (log($\frac{L_{2-10 keV}}{{\rm erg\,s}^{-1}}$)~>~42.9) AGN of \citet{Mullaney2011}, also normalized at 30 $\mu$m. }
  \label{espectro-candidatos}
\end{figure}

  In summary, for our sample of 33 RSA Seyferts we found
  a small fraction of galaxies   (6/33, 18 per cent) with a 
  significant contribution from AGN heated dust to
  the nuclear $70\,\mu$m emission. This fraction is similar to the findings of
  \cite{Mullaney2011} for a sample of X-ray selected sample with no evidence of host emission in
  their MIR {\it Spitzer}/IRS spectra. They only identified 4 galaxies out  of the 25 AGN with FIR emission
  dominated by the AGN. This demonstrates that our FIR method
  to select  galaxies with significant
  nuclear  $70\,\mu$m AGN emission for optically selected
  Seyferts produces similar results to their MIR based method for X-ray selected AGN.

\section{Summary and conclusions}
\label{Section-conclusions}

In this work we  studied the nuclear and integrated FIR ($70-500\,\mu$m)
emission of 33 nearby (median distance of 30\,Mpc) Seyfert
galaxies from the RSA catalogue using
{\it Herschel}/PACS and SPIRE imaging observations.
 We selected these galaxies because they
are nearby and have estimates of their nuclear and integrated SFR. The goal was to
  identify galaxies with a significant fraction of their $70\,\mu$m emission
  produced by dust heated by the AGN  by taking advantage  of the
  broad FIR spectral
  coverage and the good angular resolution at $70\,\mu$m
  (median 0.8\,kpc FWHM for our sample). We analysed
the FIR properties of our sample, such as the unresolved 70 $\mu$m emission
and the nuclear ($r=1\,$kpc and $r=2\,$kpc) and integrated
FIR colours.  We fitted  grey-bodies to their SEDs and derived
nuclear and integrated SFR.  We finally put forward four criteria
  to select galaxies 
  whose nuclear $70\,\mu$m emission has a significant AGN contribution.
  These were: (1) elevated $f_\nu(70\,\mu{\rm m})/f_\nu(160\,\mu{\rm m})$) flux ratios
  to the typical colours of star forming galaxies, (2)  dust
  temperature gradient higher than typical values of star forming galaxies, (3)
  $70\,\mu$m excess emission with respect to the fit of the integrated
  FIR SEDs with a grey body with a fixed dust emissivity $\beta=2$, and (4)
 excess of nuclear SFR obtained from 
$70\,\mu$m over SFR from mid-infrared indicators.
The main results are as follows.

\begin{itemize}
 \item

    At 70 $\mu$m most RSA Seyfert galaxies  (85 per cent) in our sample have a nuclear $r=2\,$kpc
   contribution to the total flux greater than 50 per cent. This is in good agreement with
   results for the {\it Swift}/BAT AGN sample of \cite{Mushotzky}. 
     The derived  $70\,\mu$m sizes (FWHM) indicate that a significant
     fraction of this
     emission arises from regions $0.3-2\,$kpc in size.

\item

The nuclear $f_\nu(70~\mu m)/f_\nu(160~\mu m)$ flux ratio is 
higher in the nuclear regions than for the entire galaxy.
The integrated $f_\nu(70~\mu m)/f_\nu(160~\mu m)$ flux ratio distribution is 
statistically indistinguishable from the distribution of the {\it Swift}/BAT AGN
sample of \cite{Melendez} and from the normal galaxies of the KINGFISH sample \cite{Dale}.
 This confirms previous results that the integrated FIR emission of Seyfert galaxies is 
 in general dominated by the emission from their host galaxies rather than from the AGN.

\item

The grey body fits to the nuclear regions and
the total galaxy SEDs show that for a given galaxy the nuclear regions ($r = 1\,$kpc)
have the highest temperatures, in agreement 
with the spatially resolved maps of the dust temperatures of nearby Seyfert and
normal galaxies \citep[see e.g.][]{Bendo,SanchezPortal2013}.  The fitted
  dust  temperatures ($21 < T < 38\,$K) and dust emissivities ($1.0 < \beta < 2.0$)
  from the integrated SEDs are
  similar to those of normal galaxies.

\item
   When fitting the integrated SEDs using $\beta$=2, which is the
    typical value for star forming galaxies \citep{Li-Draine}, we found that
    7 Seyfert galaxies in our sample have a $70\,\mu$m excess over the fit greater
    than 28 per cent.   This suggests the presence of a hotter dust component, which could
      be associated to the presence of a nuclear starburst and/or dust heated by the AGN.

\item
The $70\,\mu$m nuclear SFRs derived with the $C_{70\mu {\rm m,galaxy}}$
coefficient are on average 4 times higher than those obtained
by  DSR2012 using the $11.3\,\mu$m PAH feature. 
 \cite{Mushotzky} and \cite{Petric} also 
found a discrepancy of 3 between the SFR calculated through 70 $\mu$m and 
through $11.3\,\mu$m PAH feature.

\item Taking into account the four criteria defined to
  select galaxies 
  whose nuclear $70\,\mu$m emission has a significant AGN contribution,
we found that   16 galaxies (48 per cent of the sample) satisfy at least 
one of these criteria, whereas  10 satisfy half or more of the criteria.

\item  By careful examination of the  10 candidates
  satisfying at least half of the criteria, we selected  six 
  RSA Seyfert galaxies  (18 per cent of the initial sample)
  whose nuclear ($r=1-2\,$kpc) $70\,\mu$m emission has
  a significant ($\sim 40-70$ per cent) contribution from  dust heated by the AGN.
  These galaxies are IC~5063, NGC~3783, NGC~4151, NGC~5347, NGC~7213, and NGC~7479.
  Four of them are Sy1 and 2 of them are Sy2. None of
  them show  $11.3\,\mu$m PAH emission on scales of
  tens of parsecs from high angular resolution MIR spectroscopy or high nuclear SFR.

  \item
  Our FIR method
  to select 
  galaxies whose nuclear $70\,\mu$m emission has a significant AGN contribution
  for optically selected
  Seyferts produces similar results to the \cite{Mullaney2011} 
  MIR based method for X-ray selected AGN, in terms of the 
  fraction of galaxies dominated by the AGN at 70 $\micron$ 
  and the spectral shapes between 5 and 70 $\micron$.

\end{itemize}

 The criteria defined in this work provide a good way to select statistically
 Seyferts with significant contribution of the AGN at 70 $\micron$ using Herschel data.

\subsection*{Acknowledgements}

We thank the referee for valuable comments that helped improve
the paper.

 J.G.-G. acknowledges financial support from the Universidad de 
Cantabria through the Programa de Personal
Investigador en Formaci\'{o}n Predoctoral de la Universidad de 
Cantabria. 
 J.G.-G., A.A.-H. and
A.H.-C. acknowledge financial support from the Spanish Ministry of Economy and 
Competitiveness through grant AYA2012-31447, which is party funded by the  
FEDER program.  P.E. acknowledges support from the Spanish Programa Nacional de
Astronom\'{\i}a y Astrof\'{\i}sica under grant AYA2012-31277. C.R.-A. is supported by 
a Marie Curie Intra European Fellowship within the 7th European 
Community Framework Programme (PIEF-GA- 2012-327934).
T.D.-S. acknowledges support from ALMA-CONICYT project
31130005 and FONDECYT project 1151239.
J.G.-G. thanks A. Khan-Ali for his support and help with \textsc{sherpa}. J.G.-G. 
also thanks A. Marcos-Caballero for his comments on statistics. The authors 
also thank E. Hatziminaoglou for her insightful comments.

Based on observations made with {{\it Herschel}, which is an
ESA space observatory with science instruments provided by European-led 
Principal Investigator consortia and with important participation from NASA.
PACS has been developed by a consortium of institutes led by MPE (Germany) 
and including UVIE (Austria); KU Leuven, CSL, IMEC (Belgium); CEA, LAM (France);
MPIA (Germany); INAF-IFSI/OAA/OAP/OAT, LENS, SISSA (Italy); IAC (Spain).
This development has been supported by the funding agencies BMVIT 
(Austria), ESA-PRODEX (Belgium), CEA/CNES (France), DLR (Germany),
ASI/INAF (Italy), and CICYT/MCYT (Spain).SPIRE has been developed by
a consortium of institutes led by Cardiff University (UK) and 
including Univ. Lethbridge (Canada); NAOC (China); CEA, LAM (France); 
IFSI, Univ. Padua (Italy); IAC (Spain); Stockholm Observatory (Sweden);
Imperial College London, RAL, UCL-MSSL, UKATC, Univ. Sussex (UK);
and Caltech, JPL, NHSC, Univ. Colorado (USA). This development has
been supported by national funding agencies: CSA (Canada); NAOC 
(China); CEA, CNES, CNRS (France); ASI (Italy); MCINN (Spain); 
SNSB (Sweden); STFC, UKSA (UK); and NASA (USA).

HIPE is a joint development by the Herschel
Science Ground Segment Consortium, consisting of ESA, the NASA Herschel 
Science Center, and the HIFI, PACS and SPIRE consortia.

This research has made use of the NASA/IPAC Extragalactic Database (NED) 
which is operated by the Jet Propulsion Laboratory, California Institute 
of Technology, under contract with the National Aeronautics and Space 
Administration. 

This research has made use of the TOPCAT software
(http://www.starlink.ac.uk/topcat/) and its tools.

\bsp	
\label{lastpage}
\end{document}